\documentclass[longauth,pdfpagelabels=false]{aa}  

\usepackage{float}
\usepackage{graphicx}
\usepackage{amsmath}
\usepackage{txfonts}
\usepackage{comment}
\usepackage[utf8]{inputenc}
\usepackage{enumitem}
\usepackage{pdfpages}
\usepackage{rotating}
\usepackage{lscape}
\usepackage{placeins}
\usepackage{orcidlink}
\usepackage{hyperref}	
\hypersetup{colorlinks=true, linkcolor=blue, citecolor=blue, filecolor=blue, urlcolor=blue}

\newcommand{\msun}        {\mbox{\rm M$_\odot$}}
\newcommand{\msunperpcsq} {\mbox{\rm M$_\odot$~pc$^{-2}$}}

\newcommand{\msunperyr}   {\mbox{\rm M$_\odot$~yr$^{-1}$}}

\newcommand{\Ico}         {\ensuremath{I_{\rm CO}}}

\newcommand{\Mmol}        {\mbox{${\rm M}_{\rm mol}$}}

\newcommand{\Reff}{\mbox{\rm R$_{\rm eq}$}}

\newcommand{\aco}{\ensuremath{\alpha_\mathrm{CO}}}

\newcommand{\ipah}{\ensuremath{I^\mathrm{PAH}_{\rm F770W}}}

\newcommand{\mstar}{\ensuremath{\mathrm{M}_\star}}

\newcommand{\sigmol}{\mbox{$\Sigma_{\rm mol}$}}
\newcommand{\sigsfr}{\mbox{$\Sigma_{\rm SFR}$}}

\newcommand{\tdep}{\ensuremath{T_{\rm dep}}}
\newcommand{\tff}{\ensuremath{T_{\rm ff}}}
\newcommand{\torb}{\ensuremath{T_{\rm orb}}}
\newcommand{\tshear}{\ensuremath{T_{\rm shear}}}

\defcitealias{Colombo_2015}{C15}
\defcitealias{Rosolowsky_2006}{R06}
\defcitealias{Bazzi_2026}{B26}

\begin{document}

    \title{The lifetime of 100,000 molecular clouds in the nearby Universe}
    \subtitle{}

    \author{
    Z.~Bazzi \inst{\ref{ubonn}}\orcidlink{0009-0001-1221-0975} \and
    M.~I.~N.~Kobayashi \inst{\ref{Nifs},\ref{cologne}}\orcidlink{0000-0003-3990-1204}\and
    D.~Colombo \inst{\ref{ubonn}}\orcidlink{0000-0001-6498-2945}\and
    F.~Bigiel \inst{\ref{ubonn}}\orcidlink{0000-0003-0166-9745}\and
    A.~K.~Leroy \inst{\ref{ohio}, \ref{columbus}}\orcidlink{0000-0002-2545-1700} \and
    S.~E.~Meidt \inst{\ref{gent}}\orcidlink{0000-0002-6118-4048} \and
    R.~S.~Klessen \inst{\ref{ita},\ref{IWR}}\orcidlink{0000-0002-0560-3172} \and
    E.~Rosolowsky \inst{\ref{alberta}}\orcidlink{0000-0002-5204-2259} \and
    R.~Chown \inst{\ref{algoma}}\orcidlink{0000-0001-8241-7704} \and
    D.~A.~Dale \inst{\ref{wyo}}\orcidlink{0000-0002-5782-9093} \and
    S. Dlamini \inst{\ref{UCT}}\orcidlink{0000-0002-2885-6172} \and
    M.~Greve \inst{\ref{ubonn}}\orcidlink{0009-0006-6310-2583}\and 
    S. K. Stuber \inst{\ref{NAOJ},\ref{MPIfR}}\orcidlink{0000-0002-9333-387X} \and
    M.~Boquien \inst{\ref{unica}}\orcidlink{0000-0003-0946-6176} \and
    T.~G.~Williams \inst{\ref{JBCA}}\orcidlink{0000-0002-0012-2142} \and 
    H.-A.~Pan \inst{\ref{tku}}\orcidlink{0000-0002-1370-6964} \and 
    M.~Querejeta \inst{\ref{oan}}\orcidlink{0000-0002-0472-1011} \and
    L.~Ramambason \inst{\ref{ita}}\orcidlink{0000-0002-9190-9986} \and
    A.~Romanelli \inst{\ref{ita}} \and
    T.~Saito \inst{\ref{ShizuokaU}\orcidlink{0000-0002-2501-9328}} \and
    L. E. C.~Romano \inst{\ref{USM}, \ref{eso}, \ref{ORIGINS}}\orcidlink{0000-0001-8404-3507} \and
    M. J. Jiménez-Donaire \inst{\ref{STScI_esa}, \ref{oan}}\orcidlink{0000-0002-9165-8080} \and
    H. Kim \inst{\ref{gemini}}\orcidlink{0000-0003-4770-688X} \and
    D.~Pathak \inst{\ref{ohio}, \ref{columbus}}\orcidlink{0000-0003-2721-487X} \and
    H.~Koziol \inst{\ref{UCSD}}\orcidlink{0009-0001-5949-1524}\and
    J.~Sutter \inst{\ref{whitman}}\orcidlink{0000-0002-9183-8102} \and
    J.C.~Lee \inst{\ref{stsci},\ref{oca}}\orcidlink{0000-0002-2278-9407} \and
    the PHANGS collaboration
    }

\institute{
    \label{ubonn}{Argelander-Institut f\"ur Astronomie, University of Bonn, Auf dem H\"ugel 71, 53121 Bonn, Germany} \and
    \label{Nifs}{National Institute for Fusion Science, 322-6 Oroshi-cho, Toki, Gifu, 509-5292, Japan} \and
    \label{cologne}{I. Physikalisches Institut, Universit\"{a}t zu K\"{o}ln, Z\"{u}lpicher Str 77, D-50937 K\"{o}ln, Germany} \and
    \label{ohio} Department of Astronomy, Ohio State University, 180 W. 18th Ave, Columbus, Ohio 43210 \and
    \label{columbus} Center for Cosmology and Astroparticle Physics, 191 West Woodruff Avenue, Columbus, OH 43210, USA \and
    \label{gent}{Sterrenkundig Observatorium, Universiteit Gent, Krijgslaan 281 S9, B-9000 Gent, Belgium} \and
    \label{ita} Universit\"{a}t Heidelberg, Zentrum f\"{u}r Astronomie, Institut f\"{u}r Theoretische Astrophysik, Albert-Ueberle-Str.\ 2, 69120 Heidelberg, Germany \and
    \label{IWR} {Universit\"{a}t Heidelberg, Interdisziplin\"{a}res Zentrum f\"{u}r Wissenschaftliches Rechnen, Im Neuenheimer Feld 205, D-69120 Heidelberg, Germany} \and
    \label{alberta}{Dept. of Physics, 4-183 CCIS, University of Alberta, Edmonton, AB, T6G 2E1, Canada} \and  
    \label{tku}{Department of Physics, Tamkang University, No.151, Yingzhuan Road, Tamsui District, New Taipei City 251301, Taiwan} \and
     \label{UCT} Department of Astronomy, University of Cape Town, Rondebosch 7701, South Africa\and
    \label{algoma}{Faculty of Computer Science \& Technology, Algoma University, Sault Ste. Marie, ON, P6A 2G4, Canada}\and
    \label{wyo} Department of Physics \& Astronomy, University of Wyoming, Laramie, WY 82071, USA \and
    \label{NAOJ}{National Astronomical Observatory of Japan, 2-21-1 Osawa, Mitaka, Tokyo 181-8588, Japan}\and
    \label{MPIfR}{Max Planck Institute for Radio Astronomy, Auf dem Hügel 69, 53121 Bonn, Germany} \and
    \label{unica}{Université Côte d'Azur, Observatoire de la Côte d'Azur, CNRS, Laboratoire Lagrange, F-06000 Nice, France} \and
    \label{JBCA}{UK ALMA Regional Centre Node, Jodrell Bank Centre for Astrophysics, Department of Physics and Astronomy, The University of Manchester, Oxford Road, Manchester M13 9PL, UK} \and
    \label{oan} {Observatorio Astron{\'o}mico Nacional (IGN), C/ Alfonso XII 3, E-28014 Madrid, Spain} \and
    \label{ShizuokaU} {Faculty of Global Interdisciplinary Science and Innovation, Shizuoka University, 836 Ohya, Suruga-ku, Shizuoka 422-8529, Japan}
    \and
    \label{USM}{Universitäts-Sternwarte, Fakultät für Physik, Ludwig-Maximilians-Universität München, Scheinerstr. 1, D-81679 München, Germany}
    \and
    \label{eso}{European Southern Observatory (ESO), Karl-Schwarzschild-Stra{\ss}e 2, 85748 Garching, Germany}  
    \and
    \label{ORIGINS}{Excellence Cluster ORIGINS II, Boltzmannstr. 2, D-85748 Garching, Germany}
    \and
    \label{STScI_esa}{AURA for the European Space Agency (ESA), ESA Office, Space Telescope Science Institute, 3700 San Martin Drive, Baltimore, MD 21218, USA}
    \and
    \label{gemini}{Gemini Observatory/NSFʼs NOIRLab, 950 N. Cherry Avenue, Tucson, AZ 85726, USA}
    \and
    \label{UCSD}{Department of Astronomy and Astrophysics, University of California, San Diego, CA 92093, USA}
    \and
    \label{whitman}{Whitman College, 345 Boyer Avenue, Walla Walla, WA 99362, USA}
    \and
    \label{stsci}{Space Telescope Science Institute, 3700 San Martin Drive, Baltimore, MD 21218, USA}
    \and
    \label{oca}{Université Côte d'Azur, Observatoire de la Côte d'Azur, CNRS Laboratoire Lagrange, 06000, Nice, France}
}

      \date{Received XXX; accepted XXX}
     \date{Received XXX; accepted XXX}
    
    \abstract
        {Multiple mechanisms are proposed for the formation of giant molecular clouds (GMCs) --- from gravitational free-fall caused by self-gravity, to stellar feedback-driven gas compression. Both the galactic environment and galaxy conditions could play an additional role in enhancing the formation via their gas surface density and star formation activity. In this paper, we make use of a catalog of 108,466 GMCs identified by F770W PHANGS-JWST imaging across 66 galaxies at a homogenized resolution of 30~pc. We measure the mass spectra in various galactic regions, whose power-law slopes vary from -1.2 to -2.0. We then estimate the formation time of each cloud using a model where GMCs form from multiple feedback compression, and find that clouds with masses $\le 10^{5}$\,\msun\ form, on average, in 20~Myr, with more massive clouds ($\sim 10^{6-7}\,\msun$) taking up to 100~Myr. We also find that cloud formation proceeds most rapidly in the central regions of galaxies, with formation timescales that are typically shorter by $\sim$ $5-10$~Myr compared to galactic disks. This effect is most pronounced in central molecular zones with enhanced star formation, highlighting the role of intense massive star formation, high molecular gas surface densities, and strong supersonic compressions in accelerating cloud formation. However, star formation is generally inefficient as the cloud lifetime is $\sim 1\,\%$ of the molecular depletion time. The formation time of clouds is $\sim 0.1$ dex longer than the free-fall time. This hints that magnetic fields, stellar feedback, or other mechanisms may prolong their formation instead of immediate free-fall collapse. This indicates a longevity of massive GMCs. The GMC ages also show only limited variation with galactocentric radius in both spiral and disk galaxies, suggesting that cloud formation proceeds similarly in these galaxy types.
        } 
        

    \titlerunning{Formation and destruction of 100,000 molecular clouds in the Nearby Universe}
    
    \authorrunning{Z. Bazzi et al.} 

    \maketitle
 
\section{Introduction}\label{S:introduction}

The interstellar medium (ISM) and the physical conditions of galaxies primarily drive their evolution \citep{McKee_2007,Somerville_2015, Klessen_2016}. Gas exists in multiple phases within the ISM \citep{Field_1969,Wolfire_1995,Wolfire_2003} --- atomic gas condenses into molecular gas \citep{Hollenbach_1999,Krumholz_2009}, which then forms the massive giant molecular cloud (GMC) structures within galaxies \citep{McKee_2007,Heyer_2015}. During their gravitational collapse, stars form \citep{Shu_1987}, and subsequently ionize, destroy, or compress their surrounding gas through stellar feedback mechanisms \citep{Krumholz_2014,Dale_2015,Chevance_2020b}.

Understanding the molecular properties and lifetimes of GMCs is therefore important to pin down how galaxies evolve from being molecular-rich and star-forming to quiescent and dominated by old stars. In \cite{Bazzi_2026} (hereafter \citetalias{Bazzi_2026}), we traced the molecular phase of the ISM by using the James Webb Space Telescope (JWST) F770W band for 66 Physics at High Angular resolution in Nearby GalaxieS (PHANGS) galaxies at unprecedented sensitivity (completeness limit $\sim 2\times 10^{3}\, \msun$) and resolution (30~pc), and built the largest extragalactic GMC catalog composed of 108,466 clouds. The F770W band is mainly known for capturing a strong polycyclic aromatic hydrocarbon (PAH) feature, a stellar continuum, and continuum emission from small hot dust grains \citep{Draine_2007,whitcomb_dust}. The advantage of using PAHs is that they can trace regions where typical molecular gas tracers, such as carbon monoxide (CO) are ``dark'' (e.g., \citealt{Wolfire_2010}) and do not trace the whole molecular gas content (e.g., \citealt{Leroy_2023a, Sandstrom_2023b}). Also, the emission from PAHs shows a close link with CO emission on kiloparsec and parsec scales, revealing a strong correlation between both emissions over three orders of magnitude of intensity \citep{Regan_2004,Gao_2019,Chown_2021,Leroy_2023a,Leroy_2023,Whitcomb_2023, chown2025}. Additionally, in \citetalias{Bazzi_2026}, we found that the specific star formation rate is key to driving differences in molecular mass distributions of GMCs across galaxies and galactic environments, suggesting that star formation, galactic environment, and the physical conditions of the galaxy are crucial in shaping GMC evolution.

To further understand the factors that influence the GMC life cycle, the lifetime of GMCs and the timescales of various physical processes (e.g., self-gravity, galactic dynamics, magnetic fields) need to be constrained. Multiple efforts (e.g., \citealt{Kawamura_2009,Fukui_2010,Kruijssen_2019b,Chevance_2020a,Kim_2022, Bonne_2023,Romanelli_2025,Kim_2025}) have suggested that the lifetime of GMCs is rather short (5--30~Myr), and clouds are dispersed by stellar feedback on even shorter timescales (1--5~Myr; e.g., \citealt{Kim_2022, Ramambason_2026}). On the other hand, the molecular depletion time at galactic scales is much longer (1--4~Gyr) compared to the cloud lifetime (e.g., \citealt{Bigiel_2008}), and the shearing (and orbital time) operates on timescales greater than the lifetime of GMCs (e.g., \citealt{Sun_2022}), suggesting that clouds do not manage to convert their molecular gas content fully into stars in their lifetime and large-scale processes only play a modest role as clouds evolve. However, a more in-depth investigation of the formation of individual GMCs in different environments is required to understand the kiloparsec-scale (e.g., shear, rotation) and sub-cloud-scale (stellar feedback) factors that may impact their evolution (e.g., \citealt{Walch_2015,Kim_Ostriker_2017}).

Stellar feedback mechanisms, such as the momentum caused by stellar winds and supernovae or the coupling to the intense radiation field of massive stars (e.g., \citealt{Rahner_2017,Rahner_2019}), play a crucial role in GMC formation and destruction (e.g., \citealt{Inutsuka_2015}). The recent PHANGS-JWST images of nearby galaxies \citep{Lee_2023, chown2025} provide a fascinating picture of high column densities in regions surrounding bubbles that might be feedback driven (see \citealt{Watkins_2023, Barnes_2023}). The GMCs will eventually form new stars, which leads to their ionization and destruction (e.g., \citealt{Hosokawa_2006,Geen_2017, Kim_2017, Ali_2018, Fukushima_2021}).

In this paper, we use the giant molecular cloud mass function model (GMCMF; \citealt{Inutsuka_2015,Kobayashi2017, Kobayashi_2018}) to estimate the formation time of individual clouds across different galactic environments. This model suggests that the formation and evolution of GMCs are mainly driven by multiple expanding H{\sc ii} regions, supernovae, and radiative feedback. The GMCMF is particularly powerful because its slope directly reflects the balance between GMC formation and destruction, providing a framework to connect the cloud population to the physics that shape GMC evolution. Observations have revealed strong environmental variations: spiral arms and central regions exhibit shallower mass functions, indicative of enhanced formation of massive clouds, whereas inter-arm regions display steeper slopes, suggesting slower formation of massive clouds (e.g., \citealt{Rosolowsky_2003,Koda_2009,Colombo_2014}, see also \citealt{Mok_2020}). These trends imply that GMC assembly and dispersal operate differently depending on the local dynamical and feedback conditions.

Alternative models for molecular cloud growth employ collisional coagulation frameworks and also make use of the cloud mass distribution (e.g., \citealt{Kwan_1979,Scoville_1979, Tomisaka_1986}). Early analytic models describe GMC assembly through cloud--cloud collisions over long timescales of several $10^{2}$~Myr, with the mass distribution shaped by the balance between collisional accretion and cloud destruction associated with star formation \citep{Kwan_1979,Scoville_1979}. Extensions of this framework employ numerical cloud--cloud collision models in a spiral gravitational potential to follow the coagulation and destruction of an ensemble of molecular clouds, showing that GMC lifetimes of order $\sim40$~Myr lead to strong spiral-arm concentration, whereas longer lifetimes ($\sim10^{2}$~Myr) produce a more uniform GMC distribution \citep{Tomisaka_1986}.

Theoretical work suggests that GMCs assemble primarily through repeated episodes of supersonic compression in the warm neutral medium driven by convergent turbulent flows (e.g., \citealt{Federrath_2012, Brucy_2025}), with large-scale gravitational instabilities providing an additional mechanism for gas accumulation and cloud growth (e.g., Parker instability; see \citealt{Parker_1966}). In a magnetized ISM, shocks rarely form molecular clouds in a single event; instead, several such compressions are required, setting an effective formation timescale that depends on how frequently shocks occur and on their orientation relative to the magnetic field (e.g., \citealt{Inoue_2008,Inoue_2009,Inoue_2012, Iwasaki_2019}). This is complemented by cloud dispersal through feedback from newly formed stars on short ($1–5$~Myr) scales (e.g., \citealt{Kim_2022}).

Here, we apply the GMCMF model to the PHANGS–JWST cloud catalogs from \citetalias{Bazzi_2026} to quantify how formation timescales vary across galactic environments and across galaxies with different physical conditions. Our goal is to identify the dominant environmental factors that accelerate or hinder cloud formation, and in turn, influence the lifetime and evolution of GMCs. The layout of the paper is as follows. In Sect.~\ref{S:data}, we describe the data and catalog used in the study. In Sect.~\ref{S:MC_Props}, we introduce the molecular cloud properties used in our analysis. In Sect.~\ref{S:timescales}, we describe the GMCMF model and the different timescales that we calculate for the GMCs. In Sect.~\ref{S:Results}, we present our results and discuss them. Finally, in Sect.~\ref{S:summary}, we summarize our results and present our conclusions.

\section{Data and catalog}\label{S:data}

We analyze 66 nearby, star–forming PHANGS galaxies for which high–resolution F770W JWST imaging \citep{Lee_2023,chown2025} sensitive to PAH emission are available. The sample spans galaxies with specific star formation rates ($\mathrm{SFR}/\mstar$, where SFR is the star formation rate and \mstar\ is the stellar mass) $\gtrsim 10^{-11}\,\mathrm{yr^{-1}}$, stellar masses $\mstar \gtrsim 10^{9.5}\,{\rm M_\odot}$, moderate inclinations ($i \lesssim 70^\circ$), and distances $D \lesssim 20$~Mpc \citep{Leroy_2021}.

\subsection{JWST F770W GMC catalog}\label{ss:jwst_catalog}

In \citetalias{Bazzi_2026}, we adopted the JWST F770W band as our primary tracer of PAHs to construct a uniform GMC catalog across 66 PHANGS-JWST galaxies \citep{Lee_2023, Williams_2024, chown2025} at a homogenized resolution of 30 pc and a completeness limit of $2\times 10^{3}$~\msun. The final catalog contains 108{,}466 clouds with associated physical properties such as molecular mass surface density (\sigmol), molecular mass (\Mmol), effective radius (\Reff), and corresponding uncertainties.

The 7.7\,$\mu$m band encompasses the prominent PAH complex dominated by C--C stretching modes from largely ionized PAHs spanning a range of sizes. The observations reach a median physical resolution of $\sim 20$~pc ($16-84$th percentile: $15-25$~pc), enabling the identification of cloud-scale structures across all targets. To account for the non-negligible contribution of the stellar continuum to the F770W band via the Rayleigh--Jeans tail, we subtract this component following \citet{sutter2024}. Specifically, for Cycle~2 galaxies \citep{chown2025} we correct the F770W emission by subtracting the stellar continuum contribution (${\rm F770W}_\star$)$\, = 0.22 \times {\rm F300M}$, while for Cycle~1 galaxies \citep{Lee_2023} we apply a correction of ${\rm F770W}_\star = 0.12 \times {\rm F200W}$. In addition, the F770W band contains a dust continuum contribution. Spectral decomposition work based on Spitzer mid-infrared spectra and synthetic JWST photometry \citep{Whitcomb_2023, Dale_2025} shows that the continuum-free PAH fraction is typically $\sim 80$--$85\,\%$. Applying this method in \citetalias{Bazzi_2026}, we find similar fractions across our sample, with lower PAH contributions in regions strongly affected by feedback (e.g., H{\sc ii} regions and galactic centers), where the PAH fraction can drop to $\sim 20\,\%$ or below. 
  
We identify GMCs from these stellar-continuum-corrected F770W intensity (\ipah) maps using the \texttt{SCIMES} framework \citep{Colombo_2015}, as described in detail in \citetalias{Bazzi_2026}. The algorithm combines dendrogram segmentation performed with \texttt{Astrodendro}\footnote{https://dendrograms.readthedocs.io/en/stable/} and spectral clustering (\texttt{SpectralCloudstering} class in \texttt{SCIMES}) to isolate coherent cloud structures. The stellar-continuum–subtracted F770W maps are subsequently converted to CO intensity (\Ico) using Equation~2 from \citetalias{Bazzi_2026} (see also \citealt{chown2025}). This relation was calibrated directly on galaxy intensity maps without an explicit dust–continuum correction, even though dust emission can become increasingly important toward galaxy centers. Because our analysis employs the same observational framework for which the calibration was derived, we do not distinguish whether the underlying signal originates from PAHs or from small hot dust grains. We therefore apply no additional continuum corrections and simply note that, particularly in central regions and H{\sc ii} regions, the emission traced by F770W reflects a mix of PAH features and dust continuum.

For the present analysis, we select the subset of 83{,}990 clouds that satisfy the GMC selection criteria defined in \citetalias{Bazzi_2026}. Specifically, we require either $\ipah > 1\,{\rm MJy\,sr^{-1}}$ or $\sigmol > 4\,{\rm M_\odot\,pc^{-2}}$, as emission below these thresholds may trace the atomic gas phase rather than molecular material \citep{Leroy_2023a, chown2025}. We also include clouds in the central regions, since we aim to quantify GMC formation and destruction timescales across all galactic environments. However, GMCs in those regions tend to overlap in velocity space, and the dust and stellar continua might dominate the emission. 

\subsection{MUSE optical IFU data}\label{ss:muse}
The PHANGS-MUSE large program \citep{Emsellem_2022} provided data for 19 galaxies from PHANGS-JWST Cycle~1 with a coverage up to 2 effective radii ($\rm R_{e}$) for all targets, which is generally consistent with the footprint of JWST observations. 

The typical angular resolution of the MUSE data in this study is $\sim$0.7\arcsec, which is sufficient to isolate individual H{\sc ii} regions from each other and minimize contribution from the surrounding diffuse ionized gas \citep{Congiu_2023}. We rely on SFR measurements provided by \cite{Belfiore_2023} using dust-attenuation-corrected measurements of the Hydrogen Balmer decrement (H$\alpha$ and H$\beta$). The formalism is SFR[\msunperyr] = $\rm C_{H\alpha}L_{H\alpha, corr}[erg\, s^{-1}] = C_{H\alpha}L_{H\alpha}10^{0.4\,k_{H\alpha}\,E(B-V)}$, where $\rm L_{H\alpha, corr}$ is the attenuation-corrected H$\alpha$ luminosity (L$_{\rm H\alpha}$), $\rm k_{H\alpha}$ is the value of the reddening curve at the H$\alpha$ wavelength, and E(B--V) is the attenuation correction factor (for more details on the calculation and assumptions, see \citealt{Belfiore_2023}).

For our analysis, we project our GMC assignment maps to the same grid as the MUSE star formation rate surface density ($\Sigma_{\rm SFR}$) maps and then calculate the SFR in each cloud (${\rm SFR}_{\rm cloud}$). We only consider GMCs that are at least comparable to the MUSE beam size. However, SFR estimates based on extinction-corrected H$\alpha$ emission on cloud scales are subject to several limitations. The H$\beta$ line is often not detected at sufficient signal-to-noise in low surface-brightness or highly obscured regions (e.g., \citealt{Emsellem_2022}), which can bias or limit Balmer-decrement corrections and lead to an underestimation of the SFR in embedded GMCs. In addition, H$\alpha$-based SFRs implicitly assume continuous star formation and a fully sampled IMF over the relevant timescale, assumptions that may not be strictly valid for individual clouds and should be noted when interpreting cloud-scale SFRs. For our analysis, we calculate the SFR only for clouds within the 19 PHANGS Cycle~1 galaxies.

\section{Molecular cloud properties}\label{S:MC_Props}

In this section, we present the relevant GMC and galaxy properties used to compute the cloud timescale. A detailed explanation of other GMC quantities is given in \citetalias{Bazzi_2026}. 

Following Equation 2 in \citetalias{Bazzi_2026}, we convert \ipah\ to \Ico. The molecular mass of the cloud is then estimated as
\begin{align}
    \mathrm{M_{\rm mol}}~[\mathrm{M_{\odot}}] 
    &= \alpha_{\rm CO}\,[\mathrm{M_{\odot}(K\,km\,s^{-1}\,pc^{2})^{-1}}] 
    \times L_{\rm CO}\, [\mathrm{K\,km\,s^{-1}\,pc^{2}}], \label{E:Mass} \\
    \intertext{where}
    \alpha_{\rm CO(2-1)} 
    &\approx 4.35 \times f(Z) \times g(\Sigma_{\star}) \times R_{21}(\Sigma_{\rm SFR})^{-1}, \\
    R_{21}(\Sigma_{\rm SFR}) 
    &= 0.65 \left( \frac{\Sigma_{\rm SFR}}{0.018} \right)^{0.125}, \nonumber \\
    \intertext{and}
    L_{\rm CO} 
    &= A_{\rm pix} \sum_{i=1}^{N} F_{i}.
\end{align}
Here, $F_i$ denotes the flux of the $i$-th pixel within a cloud, expressed in units of $\mathrm{K\,km\,s^{-1}}$, and the total cloud flux is obtained by summing over all pixels belonging to the cloud. $A_{\rm pix}$ is the area of a pixel in pc$^{2}$. \aco\ is the CO-to-H$_{2}$ conversion factor following \cite{schinnerer2024} (see Appendix~\ref{a:convfact} for the impact of using another \aco\ prescription), and $L_{\rm CO}$ is the CO luminosity of the cloud in units of $\rm K\,km\,s^{-1}\, pc^{2}$. $f (Z) = (Z/Z_{\odot})^{-1.5}$ is the CO-dark factor that depends on the metallicity ($Z$) for $0.2 < Z/Z_{\odot} < 2$ (see \citealt{schinnerer2024} for further information), where $Z_{\odot}$ is the solar metallicity ($\rm 12+log(O/H) =8.69$ as per \citealt{Asplund_2009}). It is worth noting that $f (Z)$ does not take into consideration additional factors like the dust-to-metals ratio, interstellar radiation field, cosmic ray ionization rate, and the structure of the clouds themselves, which all play an important role and further add to the uncertainty of the \Mmol\ estimation (see \citealt{schinnerer2024}). The starburst emissivity factor is $g(\Sigma_{\star}) = \mathrm{max}(\Sigma_{\star}/100, 1)^{-0.25}$, where $\Sigma_{\star}$ is the stellar mass surface density in units of $\mathrm{M_{\odot}\, pc^{-2}}$. Additionally, $R_{21}(\Sigma_{\rm SFR})$ is the line ratio between CO($2-1$) and CO($1-0$) (see \citealt{Leroy_2022} and \citealt{schinnerer2024} for more information). The metallicity is approximated as a function of galactocentric radius based on the global mass-metallicity relation of \cite{Sanchez_2019}, adopting the PP04 O3N2 calibration \citep{Pettini_2004} and extrapolating the predictions to the whole PHANGS-ALMA footprint using a metallicity gradient as per \cite{Sanchez_2014} (see \citealt{Sun_2025}).

\section{GMC timescales}\label{S:timescales}

\subsection{The GMC mass function}
In the “bubble” framework of \citet{Inutsuka_2015}, GMCs form and grow through repeated, large-scale compressions of the magnetized ISM (driven by expanding H{\sc ii} regions and supernova shells) while being dispersed by stellar feedback. This picture leads to a continuity equation for the GMC mass function (GMCMF) when neglecting cloud-cloud collisions, as those collisions do not significantly modify the GMCMF evolution, especially in the lower mass range (M$_{\rm  mol} < 10^{5.5}\, \rm M_{\odot}$; see \citealt{Kobayashi2017}):
\begin{equation} \label{eq:GMCMF} 
    \frac{\partial n_{cl}}{\partial t} + \frac{\partial}{\partial m} \Bigg(n_{cl}\bigg(\frac{dm}{dt}\bigg) \Bigg) \approx -\frac{n_{cl}}{T_{d}},
\end{equation}
where $n_{cl}(m) \simeq N_{cl}/(A\times h)/m$ is the specific volumetric number density of GMCs, $N_{cl}$ is the number of GMCs within an area ($A$) of a specific region of the galaxy, $h$ is the molecular scale height of the galactic disk, $m$ is the mass of the GMC, $dt$ is the differential time, and $T_{d}$ is the characteristic self-dispersal timescale (average onset of star-formation plus destruction of cloud timescale). The typical star-formation-onset timescale ($T_{\star}$) within GMCs is assumed to be $\sim10$~Myr, as gas in the ISM typically experiences supersonic shocks from supernovae on $\sim$1~Myr timescales (e.g., \citealt{Mckee_1977}), with the effective interval between shocks ($T_{\rm exp}$) being shorter due to additional contributions from H{\sc ii} regions. Since only a small fraction of these shocks lead to successful GMC formation ($\sim$ 3~$\%$, see \citealt{Kobayashi2017}), the characteristic timescale to assemble molecular clouds from the warm neutral medium is $T_{\rm exp}/0.03 \sim 10$~Myr. During this successful cloud assembly, those shocks also trigger filamentary structures in clouds, which form massive stars \citep{Inoue_2018, Kumar_2020, Abe_2022, Maity_2024}. We therefore adopt $T_{\star} = 10$~Myr.  
Additionally, \cite{Inutsuka_2015} shows that when running a line-radiation magnetohydrodynamical simulation and including magnetic fields, once a massive star with mass larger than 30~$\rm M_{\odot}$ forms, the majority of the mass ($>10^{5}~\rm M_{\odot}$) of the GMC is destroyed within 4~Myr (see also \citealt{Hosokawa_2006,Bonne_2023}). Thus, we adopt $T_{d} = T_{\star} + 4 = 14$~Myr (see Sect.~\ref{ss:dest} for discussion on how different $T_{d}$ impact the self-growth timescale $T_{\rm f}$).

Following \citet{Kobayashi2017}, we model the mass growth of a GMC as
\begin{equation}
\frac{dm}{dt} \;=\; \frac{m}{T_{\rm f}(m)} ,
\label{eq:gmc_dm_dt}
\end{equation}
where $T_{\rm f}(m)$ is the characteristic self-growth (or ``formation'') timescale for a cloud of mass $m$. When the mass of the GMC becomes greater than the typical mass of the swept-up shell, the growth of the GMC saturates, because the dense gas that can be used to form a cloud becomes limited by the amount of total mass in the expanding shell. A truncation is therefore introduced $T_{\rm f}(m) \;=\; T_{\rm f}\,\Big[\,\big(1 + m/m_{\rm trunc}\big)^{\beta}\Big]$, with $m_{\rm trunc}$ the truncation scale ($\sim 7.7 \times 10^{6}\, \rm M_{\odot}$ for a typical Milky Way disk; see expanding shell argument in \citealt{Kobayashi2017}) and $\beta = 10$ a steepness parameter (see Appendix \ref{a:trunc} for more details). However, for most of the mass range, $T_{\rm f}(m)$ is nearly constant and equal to a fiducial value $T_{\rm f}$. Therefore, in our analysis, we assume a constant $T_{\rm f}(m) = T_{\rm f}$ and infer this value from the mass distributions of GMCs (see below). The \emph{age to reach mass $m$ by secular growth} from a seed mass $m_{\min}$ is then
\begin{align}
{T_{\rm age}(m)} \;=\; \int_{m_{\min}}^{m} \frac{T_{\rm f}}{m'}\,dm'  \label{eq:age_integral} 
\end{align}
i.e.\ exponential mass growth, $m(t)\sim m_{\min}\exp(t/T_{\rm f})$. 

Following \cite{Inutsuka_2015} and \cite{Kobayashi2017}, when we combine Eq.~\ref{eq:GMCMF} and \ref{eq:gmc_dm_dt}, we get
\begin{equation}\label{eq:GMCMF_simp}
    \frac{\partial n_{cl}}{\partial t} + \frac{\partial}{\partial m} \Bigg(n_{cl}\bigg(\frac{m}{T_{\rm f}}\bigg) \Bigg) \approx -\frac{n_{cl}}{T_{d}}.
\end{equation}
Solving this form of the GMCMF with a steady state assumption ($\partial n_{cl}/\partial t=0$) yields a steady-state solution

\begin{align}
    n_{cl}(m) = n_{0}\bigg(\frac{m}{\rm M_{\odot}}\bigg)^{\gamma}, \label{eq:sol_GMCMF} \\
    \intertext{where}
    \gamma = -1 -\frac{T_{\rm f}}{T_{d}},  \\
    T_{\rm f} = -T_{d}(\gamma + 1). \label{eq:gamma_Tf}
\end{align}

\noindent Here $\gamma$ is the power law index, $n_{0}$ is differential number density normalized at $m = 1 \, \rm M_{\odot}$. Equation~\ref{eq:gamma_Tf} implies that the slope of the GMCMF provides information about the timescale of GMCs (see Sect.~\ref{ss:powerlaw_index}). We adopt $T_{d} = 14$ Myr as previously explained, and estimate $T_{\rm f}$ from $\gamma$. 

The $\gamma$ index can be measured from the GMC mass spectra given by the following equation
\begin{equation}\label{eq:mass_spectra}
    \frac{dN}{dM} \propto M^{\gamma} \implies N(M' > M) = N_{0} \bigg[\bigg(\frac{M}{M_{0}}\bigg)^{\gamma+1} - 1 \bigg].
\end{equation}
Equation~\ref{eq:mass_spectra} refers to the solution of the integral of the GMCMF, which yields the cumulative mass spectrum. The slope of the mass spectra can be used to infer the $\gamma$ parameter. We follow a truncated power law for our functional form (see also \citealt{Williams_1997,Colombo_2014}). The maximum mass of the distribution is presented by $M_{0}$, and $N_{0}$ refers to the number of clouds with masses larger than $2^{1/(\gamma +1)}M_{0}$ (i.e., the truncation mass where the distribution deviates from a single power law).

As a practical choice, we adopt $m_{\min}= 10^{4}\,{\rm M_\odot}$ for the minimum GMC mass or completeness limit and estimate $T_{\rm f}$ following Eq.~\ref{eq:gamma_Tf} globally in each galaxy and per galactic environment (see Appendix~\ref{a:completness} for the impact of changing the completeness limit). The completeness limit set here refers to a limit where less massive clouds may not host massive stars as frequently as more massive clouds (see \citetalias{Bazzi_2026} on the minimal contribution of clouds less massive than $10^{4}\, \msun$ to star formation) and have less self-destruction by feedback from those stars. Then, we estimate the individual-cloud formation timescale following Eq.~\ref{eq:age_integral}. In Appendix~\ref{a:convfact}, we show the impact of changing the \aco\ factor on our analysis.

\subsection{Other timescales}
Besides the basic framework that is introduced above, this section provides all the different time estimates calculated for each cloud. The measurements are compiled in Table~\ref{T:big_table} for each galaxy.
\begin{enumerate}
    \item Free-fall timescale (\tff): 
    \begin{equation}\label{eq:tff}
         T_{\mathrm{ff}} = \sqrt{\frac{\rm \pi^{2} R_{3D}^{3}}{\rm 8 G M_{mol}}}.
    \end{equation}
    This timescale describes the time required for a cloud to collapse in free fall due to its own gravity of the cloud, provided that no pressure is supporting the system. Here, $\rm R_{3D}$ is the three dimensional radius of the cloud estimated as $\rm R_{3D} = \min{\bigg[R_{eq}, \bigg(R_{eq}^{2} \sqrt{\frac{h}{2\cos{\textit{i}}}}\bigg)^{1/3}\bigg]}$, and $\rm R_{eq}$ is the equivalent radius of the cloud calculated from the number of pixels within the cloud ($\rm R_{eq} = \sqrt{A_\mathrm{cloud}/\pi}$, where $\rm A_\mathrm{cloud}$ is the area of the cloud and it is directly estimated from the number of pixels within the cloud multiplied by the area of the pixel in $\rm pc^{2}$). The molecular gas disk scale height $h$ is assumed to be $100$~pc \citep{Heyer_2015} and $\frac{h}{2\cos{i}}$ would be the inclination-corrected molecular disk scale height. 
    \item Orbital time (\torb):
    \begin{equation}\label{eq:torb}
        T_{\mathrm{orb}} = \frac{2\pi}{\Omega_{\mathrm{circ}}}.
    \end{equation}
    This is the period of the orbital revolution around the center of a galaxy. To approximate the angular velocity $\Omega_{\rm circ}$, we use rotation curve measurements from \cite{Sun_2022}, derived from CO line kinematics to find $\Omega_{\rm circ}$. The measurements are based on the PHANGS-ALMA dataset \citep{Leroy_2021}, and are done in radial bin sizes of 150\,pc. We specifically adopt the “universal rotation curve” functional form suggested by \cite{Persic_1996} for the rotation curves. Based on these best-fit analytical models of the co-rotation curves and the estimated uncertainties on the model parameters, the circular velocity ($\rm V_{circ}$), $\rm \Omega_{circ}$, and the logarithmic derivative of the rotation curve $\frac{d\, \ln \rm \, V_{circ}}{d \, \ln \rm \,R_{gal}}$ (where $\rm R_{gal}$ is the galactocentric distance) is extrapolated at each radius (see \citealt{Sun_2022} for more information). Orbital timescales for the clouds are estimated for 60 galaxies where the rotation curve has been measured.
    \item Shearing time (\tshear):
    \begin{equation}\label{eq:tshear}
         T_{\mathrm{shear}} = \rm \frac{2}{\Omega_{circ}(1-\beta)}.
    \end{equation}
    Following \cite{Sun_2022}, this is the timescale for two objects to move closer/farther by a unit length azimuthally, given that they are on two circular orbits separated radially by the same unit length. The shearing timescale is estimated for clouds in 60 galaxies where the rotation curve has been measured.
    \item Depletion time (\tdep):
    \begin{equation}\label{eq:tdep}
             T_{\mathrm{dep}} = \rm \frac{M_{mol}}{SFR_{cloud}} = \frac{1}{SFE_{cloud}}.
    \end{equation}
    This timescale corresponds to the inverse of the star formation efficiency (SFE). It indicates how long it would take a cloud to deplete its molecular gas reservoir at its current SFR. The depletion time is calculated for clouds in the 19 PHANGS-JWST Cycle~1 galaxies.
\end{enumerate}

\section{Results and discussion}\label{S:Results}

\begin{figure*}[h]
    \centering
    \includegraphics[width=1\textwidth]{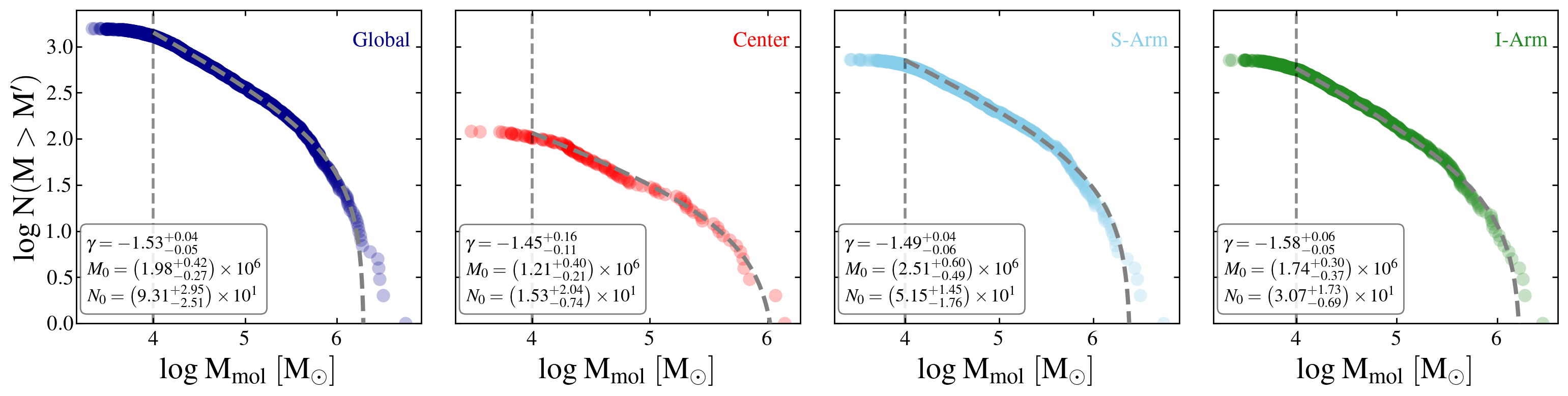}
    \caption{Example of the truncated power law fit on NGC\,0628. We present the fit on all clouds (Global) and per galactic environment (Center, S-Arm, I-Arm). The vertical dashed line is the completeness limit that we set in our analysis. The dashed curve is the truncated power law fit. The fit parameters are presented in the box in the lower left region of each plot.}
    \label{fig:mass_spectra}
\end{figure*}

In this section, we present the truncated power law fit parameters. We then compare the GMC timescale estimates per galaxy and galactic environment, and to other timescale estimate methods. We used the PHANGS “simple” galactic environment masks \citep{Querejeta_2021}, which classify regions into center (bulge or nucleus), bar (including bar ends), spiral arm, interarm, and disk (outside the bar) in galaxies lacking spiral arms (for more details see \citetalias{Bazzi_2026}). We also investigate if there is a dependence of $T_{\rm f}$ and galactocentric radius, which might hint at large-scale processes influencing the evolution of GMCs (e.g., shear or gravity). Additionally, Table~\ref{T:big_table} represents the timescale measurements in each galaxy.

\subsection{The power law index}\label{ss:powerlaw_index}

The mass spectra of GMCs provide a straightforward way to understand the relative differences between regions within and across galaxies, as well as an intuitive understanding of whether most of the mass resides in low- or high-mass clouds. Specifically, the slope of the mass spectra would reflect which population of clouds dominates the mass distribution, with shallower slopes indicating the prominence of more massive clouds (e.g., \citealt{Colombo_2014, Vieira_2024}). A traditional way of fitting the mass spectra is either by a single power law or a truncated power law (e.g., \citealt{Williams_1997}), since the mass spectrum generally steepens at high cloud masses (e.g., \citealt{Fukui_2001, Rosolowsky_2007}). The functional form of this truncated power law is presented in Eq.~\ref{eq:mass_spectra}, where the $\gamma+1$ power law index reflects the slope. The $\gamma$ parameter is also the slope of the GMCMF, which means the prominence of low- or high-mass clouds would give us information about the evolutionary track of GMCs within a galaxy, and could be used to trace the lifetime of GMCs (e.g., \citealt{Inutsuka_2015, Kobayashi2017}).

We therefore fit a truncated power law on the mass spectra across 66 galaxies and their different galactic environments. For the fits, we apply 100 bootstrap resamples while including the error on \Mmol. The fit parameters are presented in Table~\ref{tab:index}. An example of the truncated power law fit on NGC\,0628 is displayed in Fig~\ref{fig:mass_spectra}. We use a \texttt{Python}-based fitting procedure similar to the approach used in \citet{Rosolowsky_2005}\footnote{https://github.com/low-sky/idl-low-sky/blob/master/$\newline$eroslib/mspecfit.pro}. The $\gamma$ parameter varies by up to a factor of two across galaxies, environments, and within the same environment across galaxies. This is also reflected in the log-normal distribution fit in \citetalias{Bazzi_2026}, which indicates that both the galactic environment and the galactic conditions play a role in driving different mass distributions (the different factors will be discussed in Sect.~\ref{ss:gals},~\ref{ss:envs}, and \ref{ss:galrad}). Generally, the shallowest slope is for clouds within the central regions of the galaxies, followed by disks and spiral arms. This implies that those environments host more contribution from massive clouds to their mass spectrum compared to the interarm and bar regions. However, a caveat applies to the central regions, where overlapping GMCs cannot be distinguished in two-dimensional images, potentially leading to artificially shallower slopes. Additionally, individual galaxies might show variation in their trend (see Table~\ref{T:big_table}).

Table~\ref{T:correlations_gamma_selected} shows the Spearman correlation between $\gamma$ and global galactic properties and cloud properties. The strongest correlations are with cloud-scale molecular mass surface density (\sigmol) and star formation rate surface density (\sigsfr), highlighting the role of localized processes, such as shocks from massive stars and supernovae, in enhancing cloud formation (see Sect.~\ref{ss:gals} for more information). The most notable anti-correlations are with \mstar\ and SFR. This suggests that global star formation activity and galactic potential within galaxies drive steeper slopes in the mass spectrum, possibly due to the destruction of massive GMCs by feedback processes. Meanwhile, the correlation between $\gamma$ and \Mmol\ implies that more molecular gas in galaxies leads to more massive GMCs. 

\renewcommand{\arraystretch}{1.5}
\setlength{\tabcolsep}{2.5pt}
\begin{table}[h]
\caption{Truncated power-law fit parameters of the GMC mass spectrum across the galaxies.}
\centering
\begin{tabular}{ccccc}
\hline
Env. &
$N_{\rm clouds}$ &
$\gamma$ &
$M_{0}\,[M_\odot]$ &
$N_{0}$ \\
\hline
Global & 83{,}990 &
$-1.51^{+0.13}_{-0.18}$ &
$(2.25^{+2.20}_{-1.22})\times10^{6}$ &
$(1.25^{+0.99}_{-0.85})\times10^{2}$ \\

Center & 796 &
$-1.34^{+0.05}_{-0.12}$ &
$(5.01^{+33.45}_{-2.79})\times10^{6}$ &
$(7.02^{+7.71}_{-4.87})\times10^{0}$ \\

Bar & 8{,}003 &
$-1.61^{+0.21}_{-0.14}$ &
$(2.04^{+2.82}_{-1.20})\times10^{6}$ &
$(7.81^{+32.89}_{-3.62})\times10^{0}$ \\

S-Arm & 16{,}683 &
$-1.48^{+0.09}_{-0.10}$ &
$(3.47^{+1.83}_{-1.63})\times10^{6}$ &
$(5.99^{+4.20}_{-2.61})\times10^{1}$ \\

I-Arm & 27{,}132 &
$-1.59^{+0.13}_{-0.07}$ &
$(1.80^{+0.85}_{-0.93})\times10^{6}$ &
$(9.25^{+6.54}_{-5.27})\times10^{1}$ \\

Disk & 31{,}376 &
$-1.49^{+0.13}_{-0.10}$ &
$(1.84^{+2.67}_{-1.21})\times10^{6}$ &
$(1.09^{+1.13}_{-0.86})\times10^{2}$ \\
\hline
\end{tabular}
\tablefoot{Medians and quantiles are derived from 100 bootstrap resamples of $\gamma$ within each environment across 66 galaxies with quantiles that represent the ${}^{+(\mathrm{84th-50th})}_{-(\mathrm{50th-16th})}$ percentile distribution. The bootstrap takes the error into con $N_{\rm clouds}$ denotes the total number of clouds per environment.}
\label{tab:index}
\end{table}

\begin{table}[h]
\caption{Spearman correlation coefficients ($\rho$) and $p$-values between the GMC mass function slope $\gamma$ and selected galactic/cloud properties across the 66 galaxies, ordered by decreasing $|\rho|$.}
\centering
\begin{tabular}{lcc}
\hline
\multicolumn{1}{c}{Parameter} & $\rho$ & $\log_{10} p$ \\
\hline
log \mstar\ [$M_\odot$]        & $-0.55$ & $-5.65$ \\
log SFR [$\msunperyr$]                            & $-0.47$ & $-4.22$ \\
$\rm log\, M_{\rm H_I}$ [\msun]                & $-0.19$ & $-0.86$ \\
log $\rm R_e$ [kpc]                             & $-0.05$ & $-0.14$ \\
$N_{\rm clouds}$  & $-0.02$ & $-0.07$ \\
\hline
\multicolumn{1}{c}{Cloud-Scale Parameter} & $\rho$ & $\log_{10} p$ \\
\hline
$\rm log\,\Sigma_{\rm SFR}$ [$\msunperyr\,\rm pc^{-2}$]  &  $+0.90$ &     $-1.52$ \\
$\rm log\,\Sigma_{\rm mol}$ [\msunperpcsq]  &  $+0.85$ &     $- 1.31$ \\ 
$\rm log\, M_{mol,\,med}$ [\msun]  & $+0.55$ & $-5.78$ \\
$\rm log\, R_{\rm eq,\,med}$ [pc]      & $+0.22$ & $-1.07$ \\
log SFE$_{\rm\,med}$ [1/yr]                               & $+0.11$ & $-0.43$ \\
\hline
\end{tabular}
\tablefoot{Spearman rank correlations computed across $N=66$ galaxies (one global $\gamma$ value per galaxy). The median properties are represented by ``med'' in subscript, of the fit-eligible cloud sample per galaxy. The correlation between $\gamma$ and both $\Sigma_{\rm SFR}$ and \sigmol\ is estimated in increasing property bins following Fig.~\ref{fig:Timescales_bins}.}
\label{T:correlations_gamma_selected}
\end{table}

\subsection{Cloud formation drivers within galaxies}\label{ss:gals}
\begin{figure*}[h]
    \centering
    \includegraphics[width=0.49\textwidth]{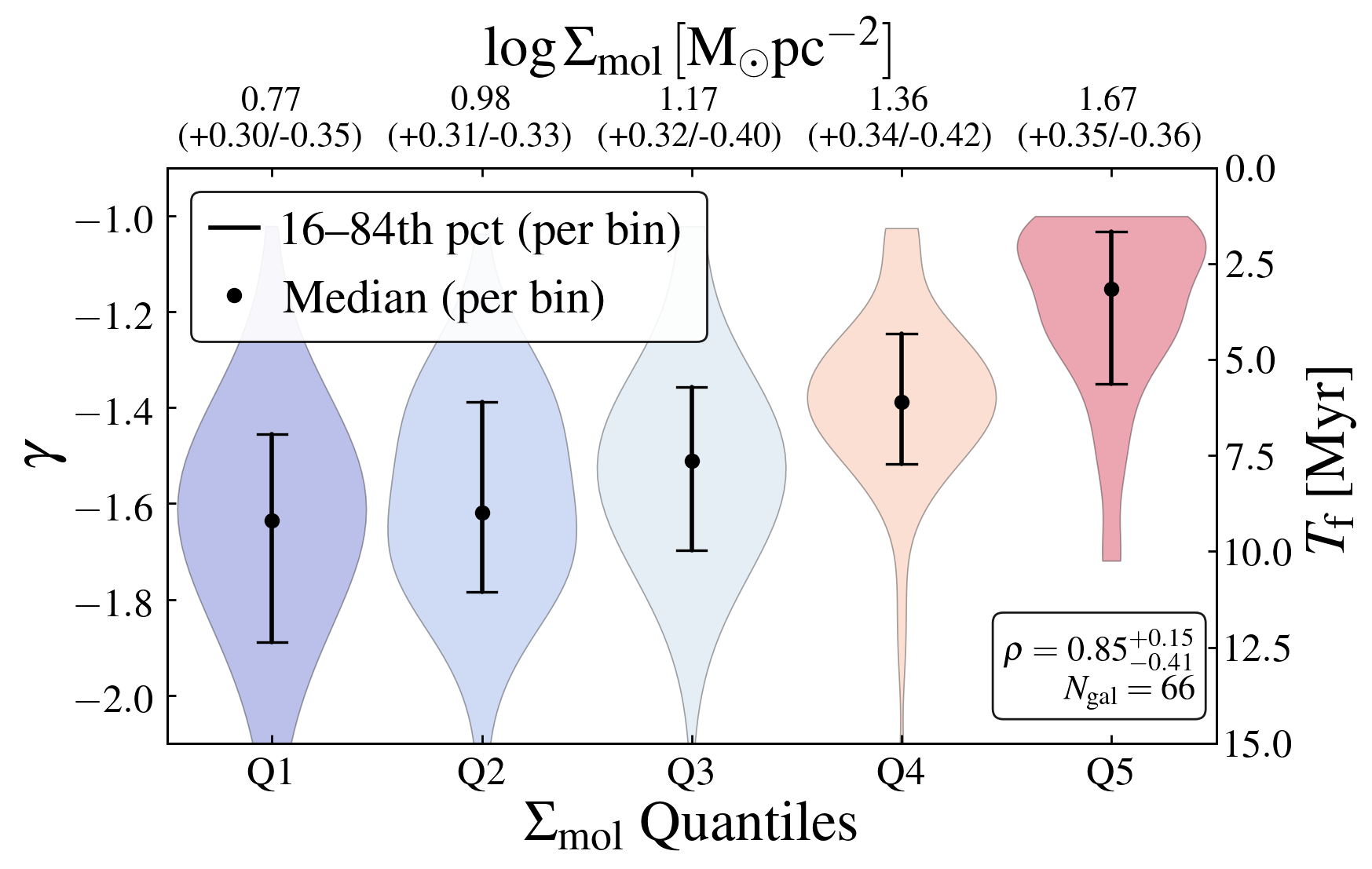}
    \includegraphics[width=0.49\textwidth]{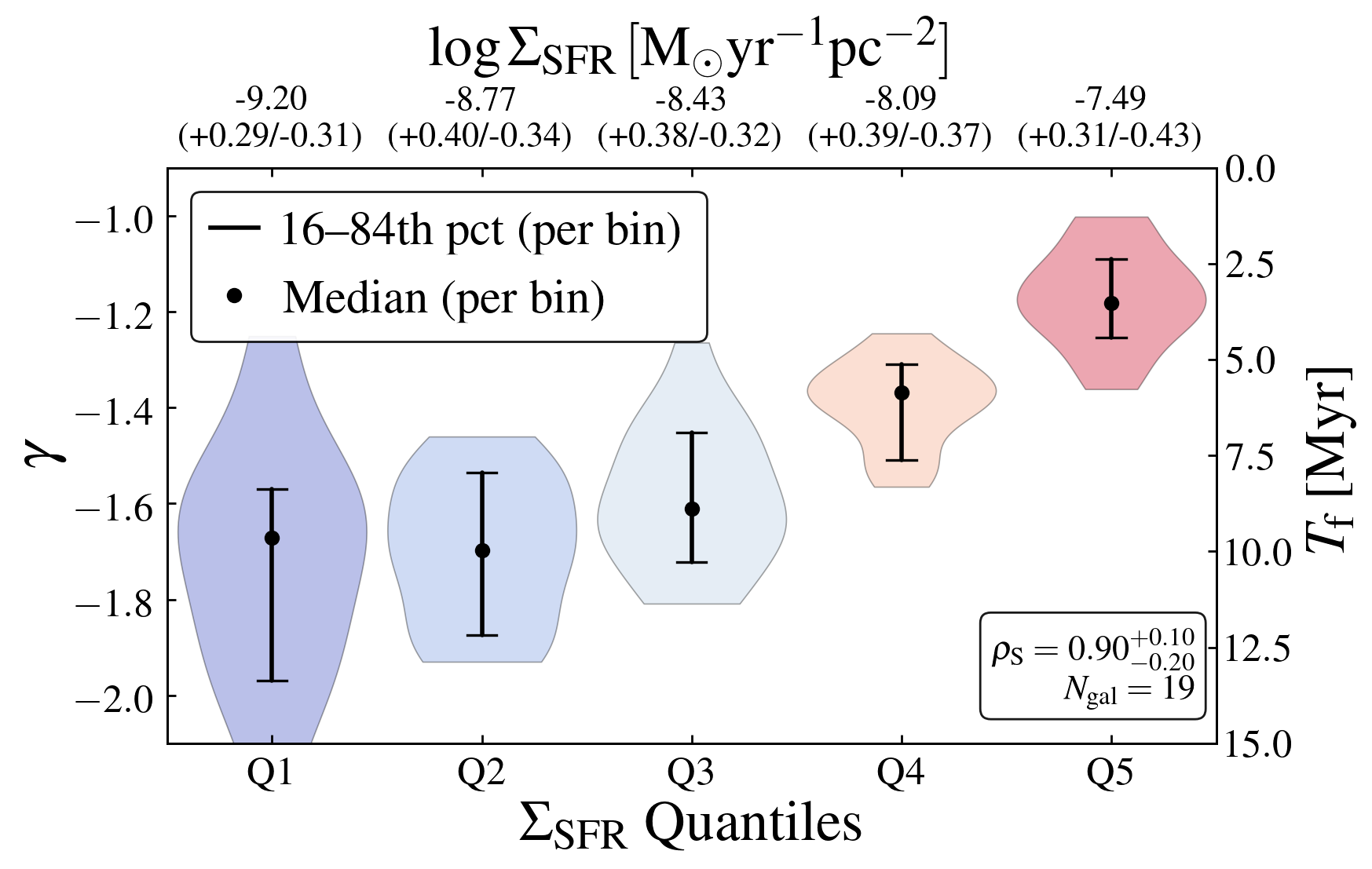}
    \caption{Violin plots showing the distribution of the GMC mass spectra slope $\gamma$ and $T_{\rm f}$ in increasing (Q1 to Q5) cloud-scale \sigmol\ (\textit{left}) and \sigsfr\ (\textit{right}) bins within 66 (PHANGS Cycle 1 and 2) and 19 (PHANGS Cycle 1) galaxies, respectively. The upper x-axis shows the property median of each bin across the galaxies, with quantiles corresponding to the ${}^{+(\mathrm{84th-50th})}_{-(\mathrm{50th-16th})}$ percentile range. The Spearman correlation coefficients ($\rho$) between $\gamma$ and $\Sigma_{\rm mol}$ (\textit{left}), $\gamma$ and $\Sigma_{\rm SFR}$ (\textit{right}), across the bins are presented in the lower right as median and quantiles corresponding to ${}^{+(\mathrm{84th-50th})}_{-(\mathrm{50th-16th})}$ percentiles range across the galaxies. Clouds in the highest \sigmol\ and \sigsfr\ bins (Q5) show on average $\sim 6.8$ and $\sim 6.9$ Myr, respectively, quicker cloud formation than the lowest bin (Q1).} 
    \label{fig:Timescales_bins}
\end{figure*}

\begin{figure*}[h]
    \centering
    \includegraphics[width=0.49\textwidth]{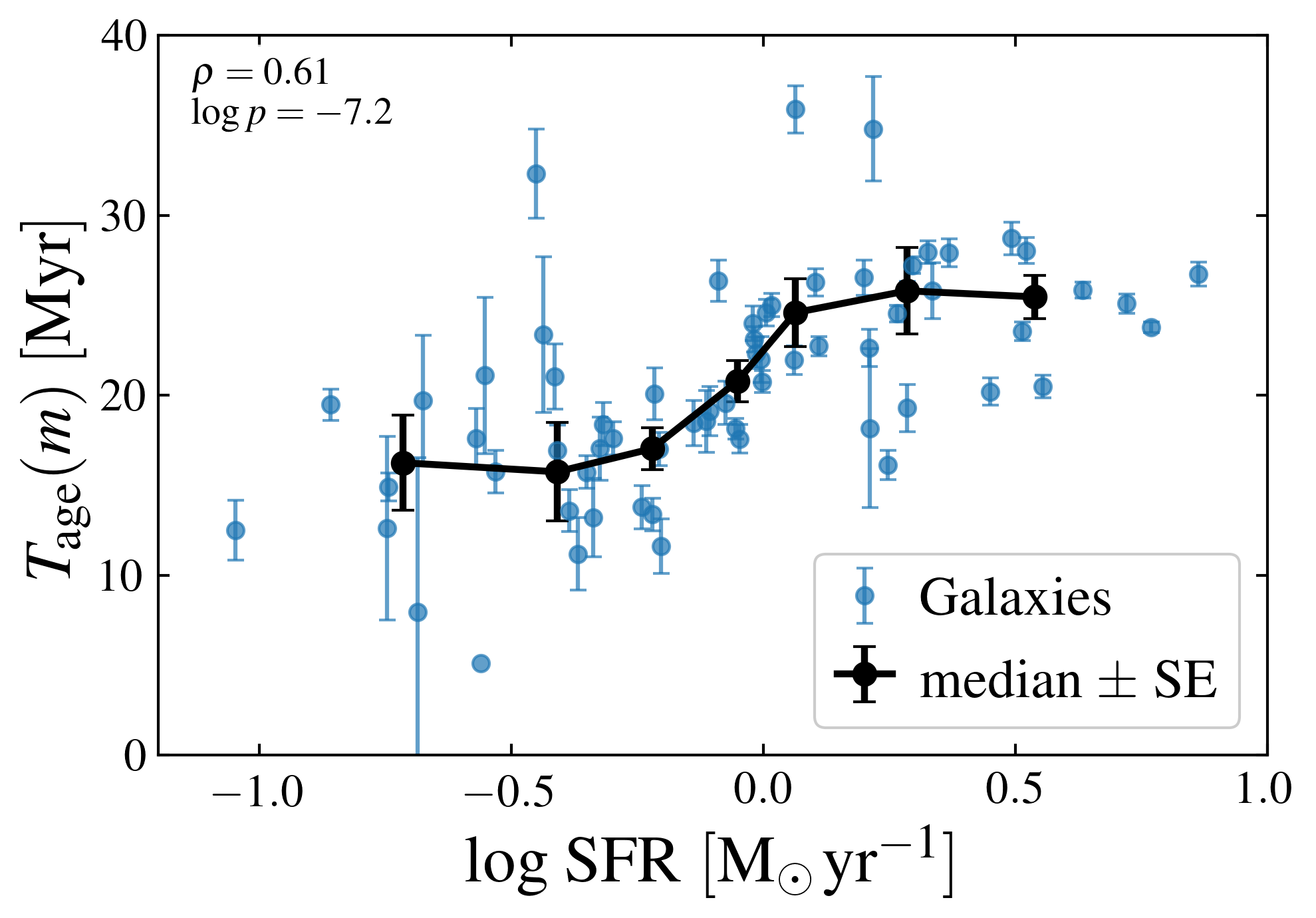}
    \includegraphics[width=0.49\textwidth]{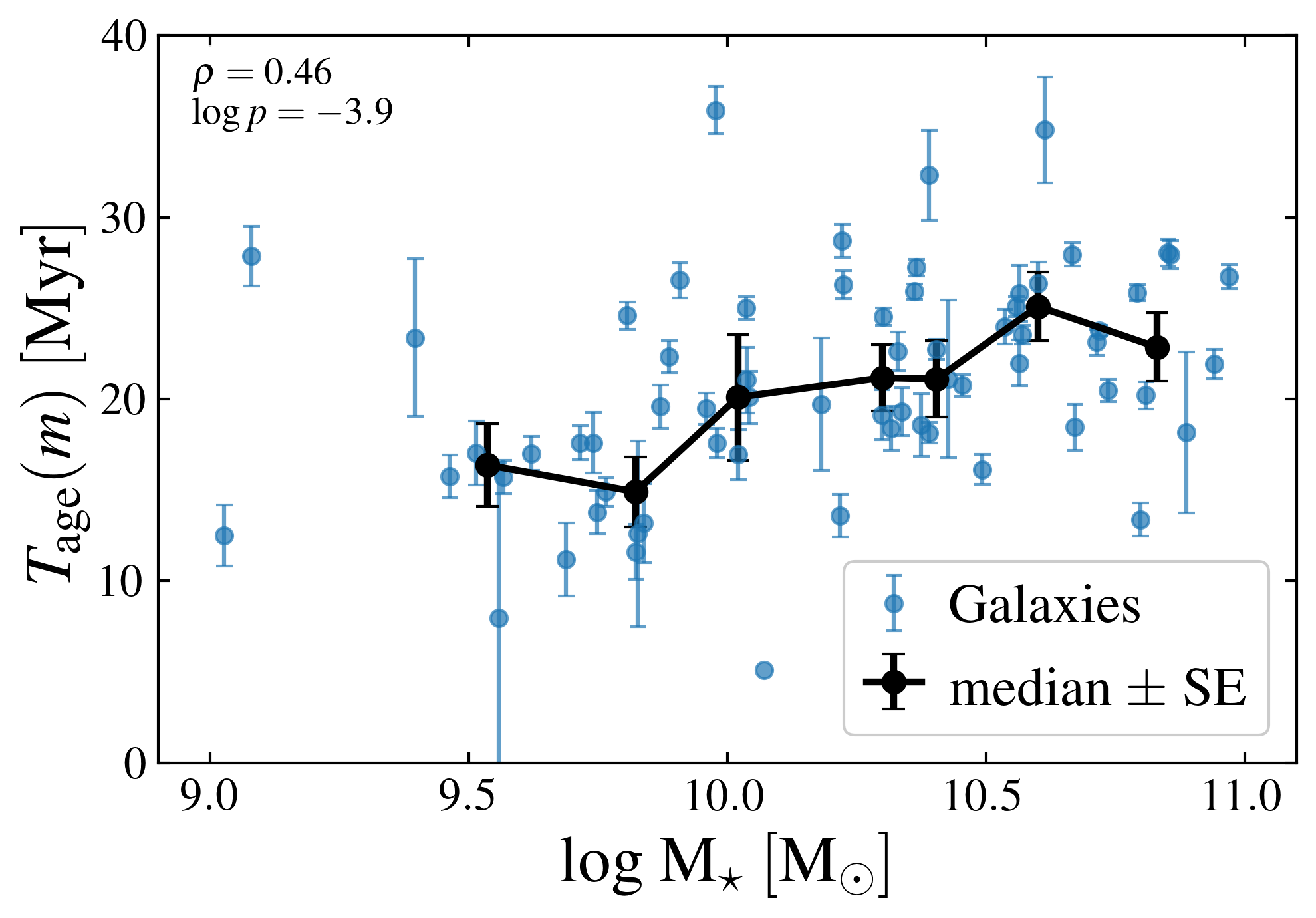}
    \caption{Scatter plots with error bars showing how the median $T_{\rm age}(m)$ varies with log SFR (\textit{left}) and log \mstar\ (\textit{right}) for the 66 galaxies in our sample. The black line is the running median with the same number of galaxies in each bin, and the error is the standard error on the median. The Spearman correlation coefficient ($\rho$) is presented in each figure with the $p$-value. Galaxies tend to have longer cloud formation time with increasing SFR and \mstar.} 
    \label{fig:SFMS}
\end{figure*}

In this section, we examine the factors that may affect the various cloud formation timescales within galaxies. For that, we test whether star formation and the gas surface densities could play a role in enhancing their formation.

In the left panel of Fig.~\ref{fig:Timescales_bins}, we bin GMCs in each galaxy into five separate bins (quantiles; Q1 -- Q5) of increasing cloud-level \sigmol, and within each galaxy, each bin has the same number of GMCs. We then compute $\gamma$ and $T_{\rm f}$ from the mass spectra in each of the bins to test whether regions with high \sigmol\ form clouds more quickly. The most notable difference is between the highest \sigmol\ quantile and the lowest one, with on average $6.8 \pm 2.2$\,Myr quicker GMC formation for the highest $\sigmol$. The highest difference is $\sim 12.1 \pm 2.8$\,Myr between Q1 and Q5, and it is for NGC\,4321 (barred spiral galaxy), $9.4 \pm 2.4$ Myr for NGC\,1566 (barred spiral galaxy), and $\sim 9.1 \pm 2.4$\,Myr for NGC\,4254 (spiral galaxy). Generally, in $\sim 73\,\%$ of the barred galaxies, there is significantly faster formation in the highest \sigmol\ quantile compared to the lowest. This might also indicate that the presence of a bar could play an additional role in accelerating cloud formation through, e.g., large-scale gas inflows. 

We also bin the galaxies in five separate bins of increasing cloud-level \sigsfr\ in the right panel of Fig.~\ref{fig:Timescales_bins}. This would help us assess whether local star formation activity, which regulates the frequency of shocks from massive stars and supernovae, has a measurable impact on cloud lifetimes. The most notable difference between the quantiles on average is Q5 and Q1 ($\sim 6.9\pm 1.9$\,Myr), and specifically in NGC\,4321 (barred spiral) with $\sim 12.0 \pm 1.9$\,Myr, NGC\,0628 (spiral galaxy) with $\sim 11.3\pm 2.2$\,Myr, and NGC\,1087 (barred disc) with $\sim 8.1 \pm 2.1$\,Myr. This hints that star formation could drive faster cloud formation through supersonic compression of the surrounding gas \citep{Kobayashi2017}. Also, the highest contrast between the \sigmol\ and \sigsfr\ bins is within spiral galaxies ($\sim 77\, \%$ of spirals), which hints that spiral arms might play an additional role in enhancing cloud formation through spiral density waves, high \sigmol\ regions, and additional supersonic compressions.

We estimate the exponential growth time $T_{\rm age}(m)$ of individual clouds assuming clouds reach their mass $m$ by secular growth from a seed mass $m_{\rm min} = 10^{4}\,\rm M_{\odot}$ using Eq.~\ref{eq:age_integral}. In Fig.~\ref{fig:SFMS}, the average age of individual clouds increases with SFR and \mstar. \cite{Kim_2022} (see also \citealt{Pan_2022}) find similar trends, although they exclude galactic centers, when estimating the cloud lifetime using CO and H$\alpha$ based analysis using the prescription of \cite{Kruijssen_2014_method}. This increase is expected, since the correlation between the molecular mass of the cloud and $T_{\rm age}(m)$ is exponential following Eq.~\ref{eq:age_integral}. This implies that more massive clouds take longer to form than smaller clouds. Given that SFR and \mstar\ are also correlated since more star-forming galaxies tend to be more massive, an increase of $T_{\rm age}(m)$ with SFR is expected. 

We further split galaxies into active (hosting an AGN) and non-active following \cite{Cetty_2010}. Out of the galaxies in the sample, 15 are classified as active. The formation timescale is similar across galactic environments between the groups. Therefore, the impact of instantaneous AGN feedback in our sample may be minimal and does not significantly reflect or influence the formation of GMCs.

\subsection{Cloud formation across galactic environments}\label{ss:envs}

\begin{figure}[h]
    \centering
    \includegraphics[width=0.49\textwidth]{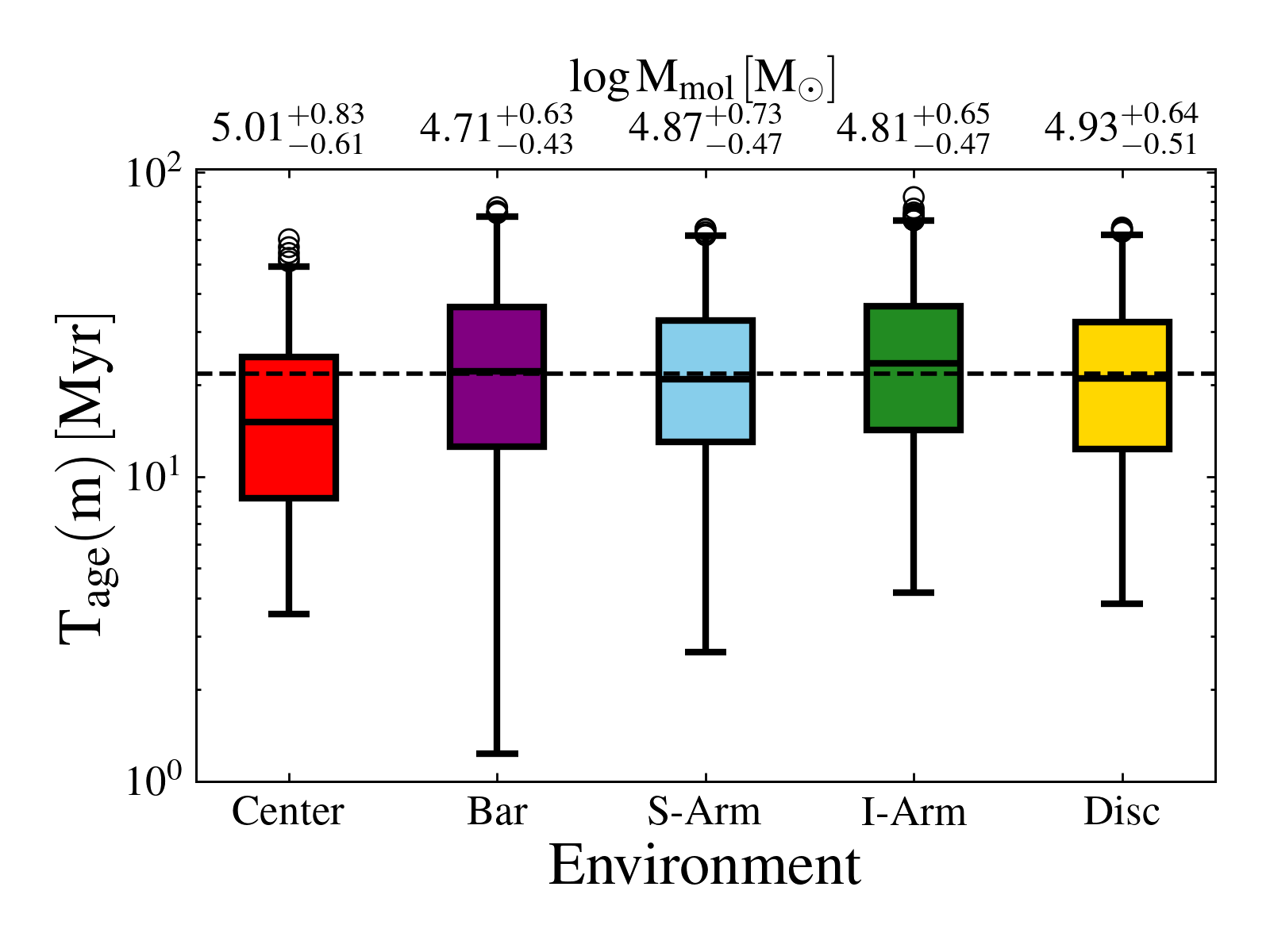}
    \caption{Bar plot with outliers (in circles) showing the exponential growth timescale, $T_{\rm age}(m)$, of GMCs across the galactic environment in 66 galaxies. The upper x-axis shows the median cloud $\Mmol$ with $\rm 84th-50th$ and $\rm50th-16th$ percentile distribution as upper and lower limits, respectively. The dashed black line represents the median timescale for all the clouds. On average, clouds with $ \Mmol \leq 10^5 \, \msun$ form in $\lesssim 20$ Myr, with more massive clouds ($10^{6-7} \, \msun$) taking up to 100 Myr. Also, central clouds form the quickest compared to other environments by $\sim 0.3$~dex. The $T_{\rm age}(m)$ distributions of each galactic environment are presented in Table~\ref{tab:gmc_T_by_env}.}
    \label{fig:T_m}
\end{figure}

As mentioned before, the dispersal time for each cloud is assumed to be mainly driven by stellar radiative feedback and is assumed to be constant (see Sect.~\ref{ss:dest} for further discussion). Meanwhile, we estimate the exponential growth time of individual clouds using Eq.~\ref{eq:age_integral}. In Fig.~\ref{fig:T_m} and Table~\ref{tab:gmc_T_by_env}, we provide median measurements across the 66 galaxies in our sample. However, it is worth noting that central regions contain reasonably higher star and dust continuum than the other environments. So there exists a caveat towards central regions, and the timescale values could be slightly higher once we subtract those contaminants.

In general, central regions exhibit the lowest $T_{\rm age}(m)$ compared to other galactic environments across the galaxies by $\sim~10-15$~Myr on average. Their extreme environments, high \sigmol, and high \sigsfr\ allow for quicker formation of clouds. Bars and interarm GMCs have, on average, $\sim 2\pm 1$~Myr longer $T_{\rm age}(m)$ than spiral arm and disk GMCs. The low $T_{\rm age}(m)$ in bar GMCs is mainly driven by low \sigmol\ and \sigsfr\ regions in dusty lanes, where the population of GMCs tends to be dominated by lower masses compared to other environments (see also \citetalias{Bazzi_2026}).

Generally, in $\sim 58~\%$ of the spiral galaxies, spiral arms have faster GMC formation compared to interarm regions. Notably, NGC\,3627, NGC\,2090, and NGC\,1672 show $9.7\pm  2.1$, $7.5 \pm 1.4$, and $6.6 \pm 1.6$\,Myr quicker cloud formation, respectively, in their spiral arms compared to interarm regions, which is mainly driven by higher \sigmol, and \sigsfr\ in arms (see \citealt{Querejeta_2024} and \citetalias{Bazzi_2026}). As highlighted in Sect.~\ref{ss:gals}, supersonic compressions and density waves within the spiral arms could play a role in quickening cloud formation in arm regions.

\renewcommand{\arraystretch}{1.5}
\setlength{\tabcolsep}{6pt}
\begin{table}[h]
\caption{Median GMC formation timescales and ages by galactic environment across the galaxies.}
\centering
\begin{tabular}{cccc}
\hline
\multicolumn{1}{c}{Env.} &
\multicolumn{1}{c}{$N$} &
\multicolumn{1}{c}{$T_{\rm f}$ (Myr)} &
\multicolumn{1}{c}{$T_{\rm age}(m)$ (Myr)} \\
\hline
Center & 796 &
$5.09^{+1.91}_{-1.79}$ &
$16.25^{+9.53}_{-6.35}$ \\

Bar & 8{,}003 &
$8.44^{+2.59}_{-2.94}$ &
$23.09^{+14.48}_{-9.66}$ \\

S-Arm & 16{,}683 &
$6.79^{+1.37}_{-1.53}$ &
$21.80^{+12.46}_{-8.29}$ \\

I-Arm & 27{,}132 &
$8.06^{+1.37}_{-1.72}$ &
$24.01^{+13.10}_{-9.61}$ \\

Disk & 31{,}376 &
$6.68^{+1.92}_{-1.97}$ &
$21.87^{+11.92}_{-8.96}$ \\
\hline
\end{tabular}
\tablefoot{
$T_{\rm f}$ values are medians of one unique formation timescale per galaxy and environment, with quantiles that represent the ${}^{+(\mathrm{84th-50th})}_{-(\mathrm{50th-16th})}$ percentile distribution across 66 galaxies. The medians and quantiles are derived from 100 bootstrap resamples that consider the error. $T_{\rm age}(m)$ medians and IQRs are computed from the cloud-level age distributions. $N$ denotes the number of clouds per environment.
}
\label{tab:gmc_T_by_env}
\end{table}
\setlength{\tabcolsep}{6pt}

\subsection{Cloud formation across galactocentric radius}\label{ss:galrad}
Examining $T_{\rm age}(m)$ as a function of galactocentric radius ($\rm R_{gal}$) provides insight into whether the average cloud formation timescale varies systematically across galaxy disks. Processes such as gravitational torques, spiral density waves, and hydrodynamic shocks (e.g., \citealt{Lin_1964, Roberts_1979, Sormani_2019, Yu_2022}) primarily act on scales larger than individual clouds, shaping the spatial arrangement and surface density of the molecular gas and driving radial gas redistribution over timescales longer than typical cloud lifetimes. By contrast, cloud formation itself may proceed through local and transient instabilities, such as Parker instability (e.g., \citealt{Parker_1966}) and swing amplification (e.g., \citealt{Goldreich_1965, Julian_1966, Toomre_1981, Fuchs_2001, Binney_2020}). The latter drives density waves, which then rapidly compress gas and form clouds on timescales comparable to the free-fall time (e.g., \citealt{Meidt_2024}). As such, trends with $\rm R_{gal}$ are generally not intended to isolate the influence of specific dynamical structures such as bars or spiral arms, which are more appropriately probed through azimuthal variations or within the galactic environment, but rather to assess whether the average cloud formation timescale exhibits systematic radial behavior.

\begin{figure}[h]
    \centering
    \includegraphics[width=0.49\textwidth]{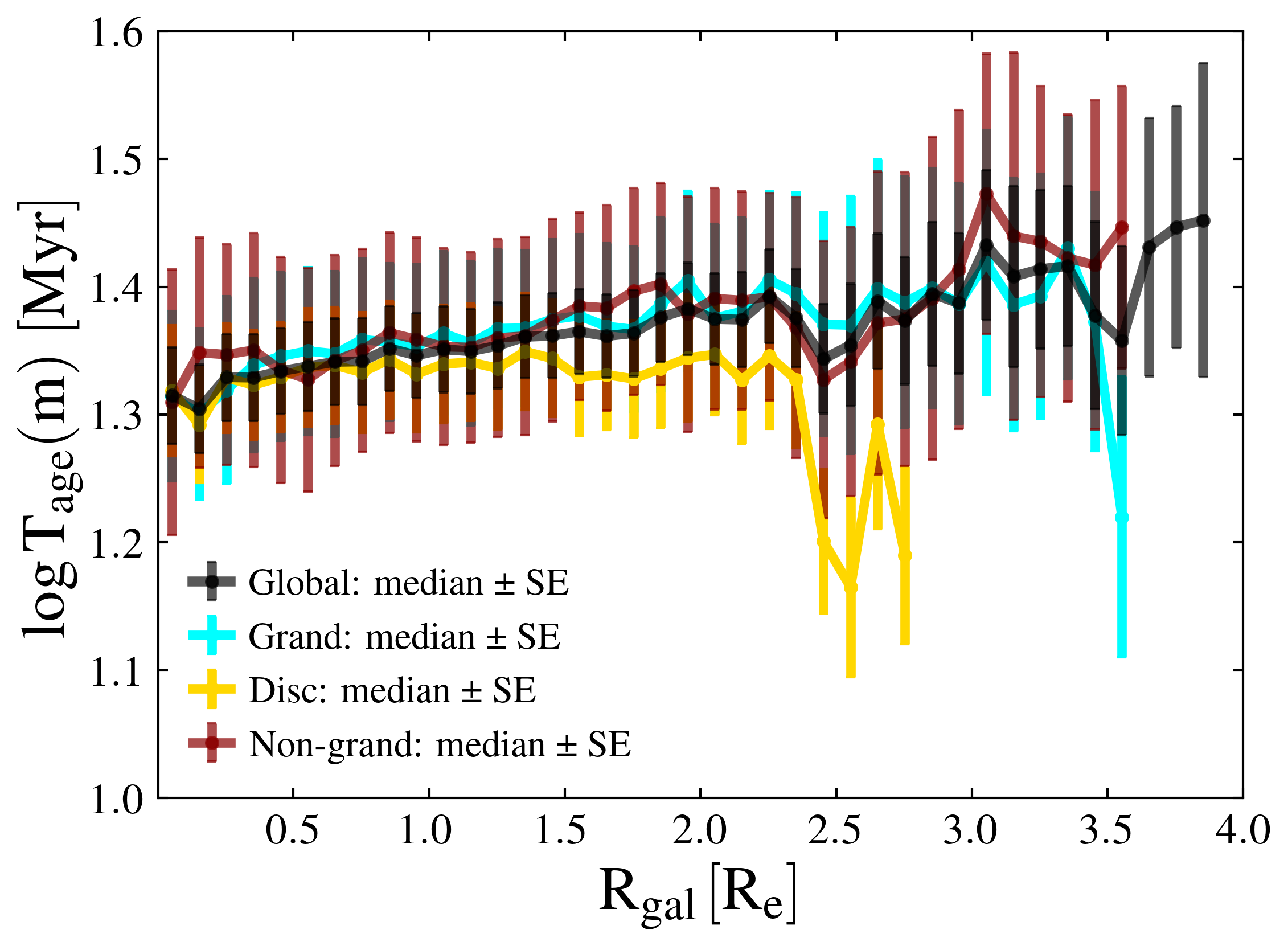}
    \caption{The growth time of the clouds, $T_{\rm age}(m)$, as a function of galactocentric radius for the sample of 66 galaxies split into grand design spirals (prominent spiral features), non-grand design spirals, and disk galaxies (no spiral features). The binned median per galaxy group is shown in the solid lines with error bars representing the standard error on the median (SE = $1.253 \times \sigma/ \sqrt{N_{gals}}$, where $\sigma$ is the standard deviation, and $N_{gals}$ is the number of galaxies in each bin). The solid black line represents the median trend for all the clouds regardless of galaxy group. All trends seem to be consistent with each other.}
    \label{fig:T_m_radial}
\end{figure}

Figure~\ref{fig:T_m_radial} shows the variation of $T_{\rm age}(m)$ as a function of $\rm R_{gal}$ for all 66 galaxies. The global trend across the galaxies shows, on average, $\sim$ 0.1 dex lower $T_{\rm age}(m)$ in the central 0.3 stellar effective radius compared to the outer regions. We then split grand-design spirals, and the rest are multi-armed, flocculent spirals or disk (lacking spiral features) galaxies following \cite{Querejeta_2024}, where galaxies are visually classified from near-infrared imaging following \cite{Buta_2015} classification scheme for galaxies in the S$^{4}$G survey. On average, $T_{\rm age}(m)$ shows no significant difference between the different galaxy groups at different $\rm R_{gal}$, indicating that cloud formation is largely alike across spiral and disk galaxies hosting similar average cloud \sigmol\ values (see \citetalias{Bazzi_2026}).

Most galaxies also exhibit flat trends in $T_{\rm age}(m)$ across all $\rm R_{gal}$. Some exceptions are galaxies hosting central molecular zones (CMZs) such as NGC 1365, NGC 1433, and NGC 3627. Those galaxies have timescales 0.2-1 dex lower towards their CMZs compared to outer regions, indicating an additional role of the extreme environment in the CMZ in driving quicker cloud formation.

\begin{figure}[h]
    \centering
    \includegraphics[width=0.49\textwidth]{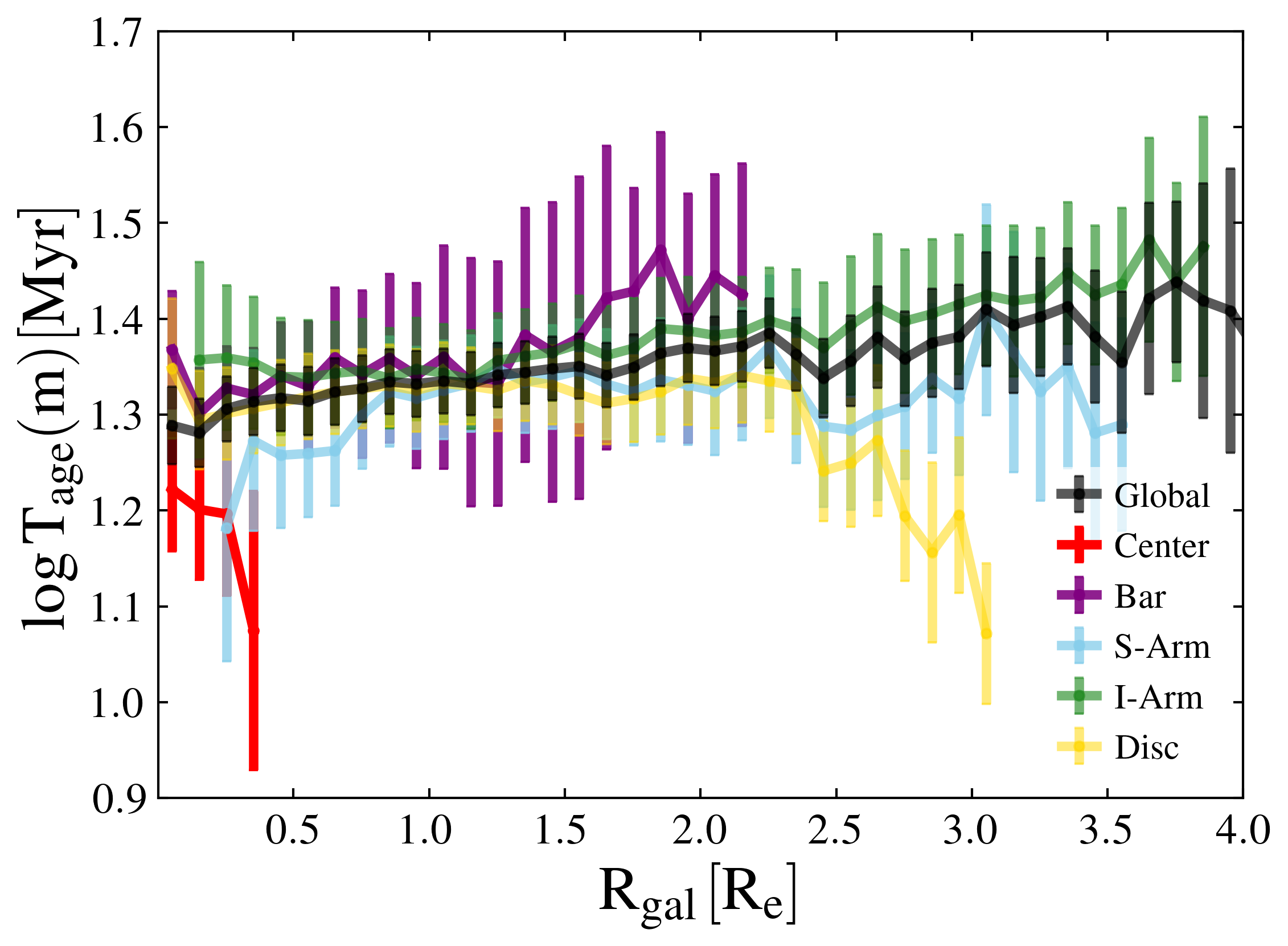}
    \caption{The growth time of the clouds, $T_{\rm age}(m)$, as a function of galactocentric radius for the sample of 66 galaxies split according to galactic environment. The binned median per galaxy group is shown in the solid lines with error bars representing the standard error on the median. The solid black line represents the median trend for all the clouds regardless of galaxy group. All trends seem to be consistent with each other, with a $\sim 0.2$ dex drop in the central regions. The scatterplot is a 2D histogram for the whole sample.}
    \label{fig:T_m_radial_envs}
\end{figure}

Figure~\ref{fig:T_m_radial_envs} shows that cloud formation timescales in spiral arm regions are, on average, slightly shorter than in interarm regions, with typical offsets of order $\sim 0.1$~dex. However, these differences are modest and comparable to the intrinsic scatter at fixed $\rm R_{gal}$. The small systematic offset suggests that spiral arms may provide enhancement to cloud formation, potentially through increased gas surface density and orbit crowding—rather than driving a fundamentally distinct formation mode. Clouds in the disk galactic environment appear to have similar timescales to those of spiral arms in the outer regions. However, towards the inner 0.3 $\rm \mathrm{R_{e}}$, spiral arms show $\sim 0.2$~dex quicker cloud formation than those in disks. This might highlight the role of large-scale inflows via arms and bars. Meanwhile, within bars, and specifically bar lanes, the cloud $\sigmol$ is relatively lower than the other environments (see \citetalias{Bazzi_2026}), which might prevent the formation of massive GMCs. It is also worth noting that within PHANGS galaxies, spiral arms have, on average, $0.1-0.2$ dex higher $\Sigma_{\star}$, $0.2-0.5$ dex higher $\sigmol$, and $0.2-0.5$ dex higher $\sigsfr$ compared to interarm regions (see \citealt{Querejeta_2024} and \citetalias{Bazzi_2026}).

\begin{figure*}[ht]
    \centering
    \includegraphics[width=1\textwidth]{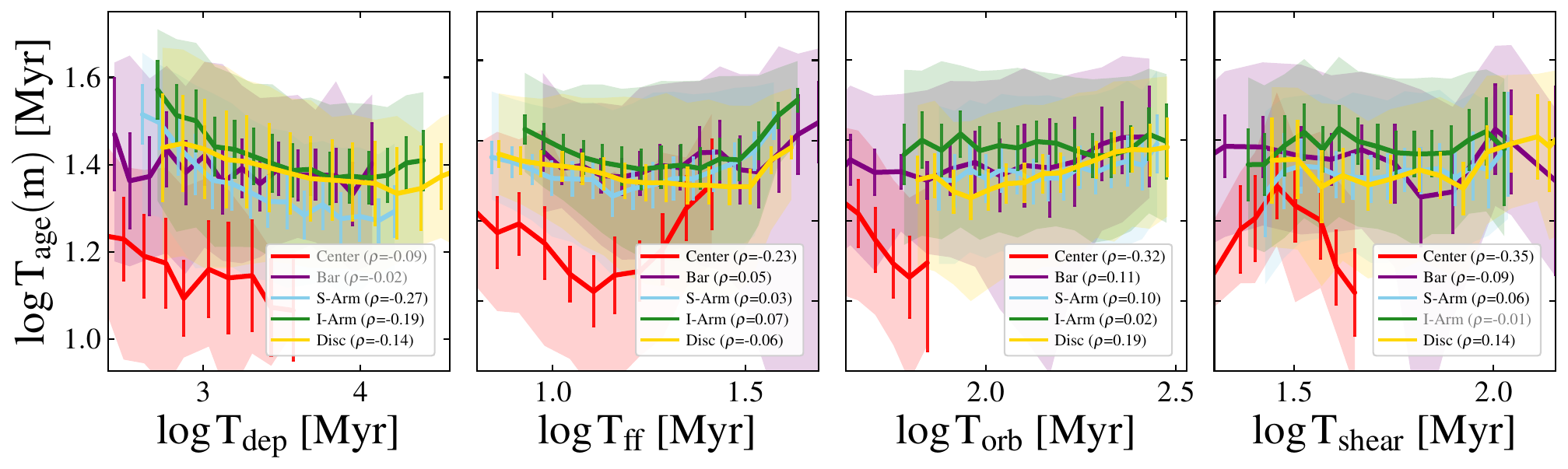}
    \caption{Growth timescale as a function of \tdep\ (Eq.~\ref{eq:tdep}), \tff\ (Eq.~\ref{eq:tff}), \torb\ (Eq.~\ref{eq:torb}), and \tshear\ (Eq.~\ref{eq:tshear}). The median trend is plotted across galactic environments for the galaxies in our sample. The shaded region represents the $\rm 84th-50th$, and $\rm50th-16th$ percentiles, and the error bar is the standard error on the median. The Spearman correlation coefficient $\rho$ is written in black when the correlation is significant ($p < 0.05$). Otherwise, it is not significant.}
    \label{fig:Timescales_corr}
\end{figure*}

\subsection{The different timescales of clouds}

Figure~\ref{fig:Timescales_corr} shows the distribution of $T_{\rm age}(m)$ with respect to all other timescales across the galaxies in our sample. The timescales presented refer to ones that take place within the cloud (\tdep, and \tff), and to those that occur on larger scales up to the scale of the galaxy (\tshear\ and \torb). \tdep\ is $\sim 1-5$~Gyr in our sample, which is consistent with other studies on molecular-dominated regions in nearby, massive, star-forming galaxies (e.g., \citealt{Bigiel_2008, Ellison_2021, Sun_2022}). \tff\ operates on timescales of $\sim 5-20$~Myr. These two timescales mean that the efficiency of star formation per free-fall time is $\sim 0.1\%-2\%$. This reaffirms that star formation is inefficient in nearby star-forming galaxies (see also, \citealt{Evans_2014, Utomo_2017, Sun_2022}). Assuming that clouds take $\sim 1-5$~Myr to disperse from stellar feedback (e.g., \citealt{Inutsuka_2015, Kim_2022, Bonne_2023}), and that the formation of individual clouds operates on timescales of tens of Myr (as depicted in Table~\ref{tab:gmc_T_by_env}), the depletion per cloud-lifetime is $\sim 1 \%$, consistent with previous studies (e.g., \citealt{Kruijssen_2019b,Chevance_2020a,Kim_2021a}), further highlighting the inefficiency of star formation.

The timescales that are associated with dynamical processes acting on kpc scales (e.g, \tshear\ and \torb) take place $\sim 60-200$~Myr. The order-of-magnitude contrast between them and \tff\ implies that the effects of galactic-scale dynamics on individual molecular clouds are minimal. The shearing time being $\sim 0.6$~dex higher than \tff\ indicates that shearing motions are generally small on cloud scales relative to motions generated by gravitational collapse (see also \citealt{Sun_2022}), and \torb\ being $\sim 0.8$~dex higher than $T_{\rm age}(m)$ indicates that molecular clouds can only last a small fraction of a complete orbital revolution around the galaxy center (e.g., \citealt{Chevance_2020a, Chevance_2020b}).

We also examine the correlation between $T_{\rm age}(m)$ and the other timescales using Spearman $\rho$ (correlation) and $p$ (probability) values. Among all comparisons, $T_{\rm age}(m)$ does not correlate with any of the timescales, as shown in Fig.~\ref{fig:Timescales_corr}. However, a typical offset of $\sim 0.1$~dex between $T_{\rm age}(m)$ and $T_{\rm ff}$, with $T_{\rm age}(m) > T_{\rm ff}$, hints that cloud lifetimes exceed the timescale required for gravitational collapse. This implies that additional physical processes, such as stellar feedback, act to slow collapse and prolong cloud evolution beyond a single free-fall time.

\subsection{Cloud destruction across galaxies}\label{ss:dest}

In the previous sections, we assumed a star-formation-onset timescale within GMCs of 10~Myr ($T_{\star} = 10$~Myr; see Sect.~\ref{S:timescales}) and a destruction time of 4~Myr once the onset occurs. Therefore, the self-dispersal timescale $T_{\rm d}$ was set to be $T_{*} +4 = 14$~Myr. The underlying oversimplified assumption of a constant cloud-destruction time is that SFE is universal for the population-average cloud properties across a mass spectrum (see \citealt{Inutsuka_2015}). However, cloud destruction might not be universal (e.g., \citealt{Federrath_2012,Wainer_2026}), and recent studies showed that the feedback time within clouds can vary from $1-5$~Myr with an average of $\sim 4$~Myr (see \citealt{Chevance_2020b, Kim_2021a, Kim_2022,Ramambason_2026}). Therefore, in this section, we test how $T_{\rm f}$ varies when changing the prescription of $T_{d}$.

Figure~\ref{fig:Tf_Td} shows that upon varying $T_{d}$ between $10+ 1$~Myr and $10 + 5$~Myr (destruction timescale varies between $1-5$ Myr across galaxies; see \citealt{Kim_2022}), the formation time increases by a factor of $\gamma$. Steeper slopes imply longer $T_{\rm f}$ values. Meanwhile, shallower slopes imply faster cloud formation. However, if stellar feedback is driving cloud destruction at longer timescales, the average formation time of clouds within galaxies would be affected by an additional $\Delta T_{\rm f} = -(\gamma +1) \times \Delta T_{d}$ according to the GMCMF prescription. Thus, the corrected difference of a varying $T_{d}$ from comparing different formation timescales across two different mass spectra, which is depicted as the shaded region in Fig.~\ref{fig:Tf_Td}, would be 
\begin{eqnarray}\label{E:Tf_corr}
    \Delta T_{\rm f_{12, corr}} = (T_{\rm f_{1}} + \Delta T_{\rm f_{1}}) - (T_{\rm f_{2}} + \Delta T_{\rm f_{2}}) \\ = \Delta T_{\rm f_{12}} - \gamma_{1}\Delta T_{d_{1}} + \gamma_{2}\Delta T_{d_{2}}\nonumber \\
    = \Delta T_{\rm f_{12}} +\epsilon. \nonumber
\end{eqnarray}
Here, both indices $\gamma_{1}$ and $\gamma_{2}$ are constants per galaxy. However, the variables are $\Delta T_{d_{1}}$ and $\Delta T_{d_{2}}$. If $T_{d}$ is universal, then $\Delta T_{\rm f_{12, corr}}$ will be equal to $\Delta T_{\rm f_{12}}$. The power law index $\gamma$ could vary between $-1.2$ and $-2.0$, and $T_{d}$ could vary between 1 and 5~Myr. Assuming $T_{d}$ is accurately calculated, with a variable SFE prescription, then the SFE-dependent factor ($\epsilon =  -\gamma_{1}\Delta T_{d_{1}} + \gamma_{2}\Delta T_{d_{2}}$) would be introduced.

It is worth noting that if $\epsilon$ changes, it creates more contrast between the galactic environments. Multiple scenarios are envisioned if $\epsilon$ changes. Some of which are: low \sigmol\ clouds could be destroyed faster than high \sigmol\ clouds, thus leading to similar $T_{\rm f}$ values between interarm and spiral arm regions. High \sigmol\ clouds could be destroyed faster than low \sigmol\ clouds, thus leading to more contrast in $T_{\rm f}$ between arms and interarms. However, our $\rm T_{age}(m)$ values, given a constant $T_{d}$, are within the range of estimates from previous works (e.g., \citealt{Chevance_2020a,Kim_2022, Kim_2025}).

\section{Summary and conclusions}\label{S:summary}

In this work, we used PAH-based GMCs identified from stellar–continuum–subtracted \textit{JWST}/F770W maps, combined with PHANGS--ALMA CO(2--1) measurements and ancillary galaxy properties, to characterize the secular growth and evolutionary timescales of molecular clouds in 66 nearby star-forming galaxies. Cloud formation is interpreted within the steady state cloud mass function framework of \citet{Inutsuka_2015} and \citet{Kobayashi2017}, which considers that GMCs form and grow through multiple repeated stellar feedback compressions. This model links the GMC mass function slope to a characteristic self-growth timescale $T_{\rm f}$ and a mass-dependent secular age $T_{\rm age}(m)$ (Eq.~\ref{eq:age_integral}). We also assume a constant GMC self-dispersal timescale of 14~Myr in our analysis. Below, we summarize our main findings.

\begin{enumerate}
    \item The slope $\gamma$ of the GMC mass spectra varies across different environments: 
    The values of $\gamma$ are generally between $-1.2$ and $-2.0$. Centers have the shallowest slopes and the highest maximum masses, followed by spiral arms and disks. This and the positive correlation of $\gamma$ with \Mmol\ hint that massive clouds dominate the mass distribution in those environments compared to bars and interarms.
    
    \item Average age of clouds vary according to secular growth $T_{\rm age}(m)$: 
    Clouds with masses $\leq 10^5$~\msun\ form in $\sim 20$~Myr, while more massive clouds ($10^{6-7}$~\msun) take up to $100$~Myr.
    
    \item Cloud formation varies systematically across galaxies: 
    Cloud growth proceeds fastest in systems with higher molecular gas surface densities and more massive GMC populations, indicating that dense, gas-rich conditions favor efficient secular cloud assembly.

    \item High-\sigmol\ and -\sigsfr\ regions accelerate cloud growth: 
    Within individual galaxies, clouds in regions of high-\sigmol\ or -\sigsfr\ display shorter $T_{\rm f}$ than those in low-\sigmol\ or -\sigsfr\ environments. The contrast can reach several Myr in barred, actively star-forming systems. This suggests that the gas density, stellar feedback, dynamical compression, or all together could promote faster GMC growth.

    \item Environmental dependence of $T_{\rm age}(m)$. Median cloud ages differ substantially between galactic environments: 
    Central regions show the shortest $T_{\rm age}(m)$ (typically $\sim 16$~Myr), $\sim 5-10$~Myr lower than in spiral arms, inter-arms, or disks. Interarm and bar environments show the longest $T_{\rm age}(m)$, which is mainly driven by low \sigmol\ within the disk and bar lanes. 

    \item Radial gradients reflect the influence of central molecular zones: 
    Across the full sample, $T_{\rm age}(m)$ decreases toward small galactocentric radii, with a typical drop of $\sim 0.1 -0.2$\,dex inside $\sim 0.3\,\rm \mathrm{R_{e}}$. Galaxies hosting a bright, dense central molecular zone exhibit the strongest gradients, with central clouds forming up to $\sim 0.2 - 1.0$~dex faster than those in their outer disks. In contrast, galaxies lacking prominent non-axisymmetric structure show nearly flat radial trends.

    \item Hierarchy of internal and galactic dynamical timescales: 
    The GMC depletion time operates at $\sim$Gyr scales, while free-fall times are $\sim 5-20$~Myr. Shear and orbital times are much longer ($\sim 60-200$~Myr). The strong contrast between these quantities demonstrates that GMC evolution unfolds on timescales much shorter than those associated with galactic rotation or shear. However, shearing processes such as swing-amplification might still play a role in cloud formation.

    \item Cloud growth and free-fall time: 
    $T_{\rm age}(m)$ is on average $\sim 0.1$ dex higher than \tff\ (see Fig.~\ref{fig:Timescales_corr}), which hints that clouds might persist for several free-fall times, requiring the presence of non-gravitational support such as turbulence, magnetic fields, or stellar feedback.

    \item Implications for GMC lifecycles and star formation efficiency: 
    Combining $T_{\rm age}(m)$, \tff, and plausible dispersal times due to feedback (a few Myr), GMCs are expected to convert only $\sim 1\%$ of their gas into stars over their lifetime. This is consistent with a picture in which cloud assembly, collapse, and destruction are jointly regulated by self-gravity, feedback, and large-scale galactic flows, leading to inherently inefficient star formation in nearby disk galaxies.
\end{enumerate}

In summary, we find cloud lifetimes, on average, consistent with previous efforts that use cloud identification methods \citep{Blitz_2007, Fukui_2008, Meidt_2015, Corbelli_2017}, statistics of sight line fractions with only CO or only H$\alpha$ or both types of emission \citep{Schinnerer_2019, Pan_2022}, and those using that use the tuning fork analysis \citep{Kruijssen_2019b, Chevance_2020b, Kim_2021a, Kim_2022, Kim_2025}. Future work would include a more focused analysis to pin down the destruction timescale of clouds. Given that this timescale varies per galaxy (e.g., \citealt{Kim_2022}), and could also vary within each galaxy, a prescription of $T_{d}$ that takes that into account would be beneficial in understanding the mass distribution in galaxies.

\begin{acknowledgements}
This work has been carried out as part of the PHANGS collaboration. This work is based on observations made with the NASA/ESA/CSA JWST. The data were obtained from the Mikulski Archive for Space Telescopes at the Space Telescope Science Institute, which is operated by the Association of Universities for Research in Astronomy, Inc., under NASA contract NAS 5-03127 for JWST. These observations are associated with programs 2107 and 3707.
ZB, DC, and FB gratefully acknowledge the Collaborative Research Center 1601 (SFB 1601 sub-project B3) funded by the Deutsche Forschungsgemeinschaft (DFG, German Research Foundation) – 500700252. M.I.N.K is supported by Grants-in-Aid from the Ministry of Education, Culture, Sports, Science, and Technology of Japan (JP22K14080). S.K.S is supported by an International Research Fellowship of the Japan Society for the Promotion of Science (JSPS).
MB acknowledges support from the ANID BASAL project FB210003. This work was supported by the French government through the France 2030 investment plan managed by the National Research Agency (ANR), as part of the Initiative of Excellence of Université Côte d’Azur under reference number ANR-15-IDEX-01.
HAP acknowledges support from the National Science and Technology Council of Taiwan under grant 113-2112-M-032-014-MY3.
LECR is supported by the Deutsche Forschungsgemeinschaft (DFG, German Research Foundation) under Germany´s Excellence Strategy – EXC 2094/2 – 390783311.
MQ and MJJD acknowledge support from the Spanish grant PID2022-138560NB-I00, funded by MCIN/AEI/10.13039/501100011033/FEDER, EU.
RSK acknowledges financial support from the ERC via Synergy Grant ``ECOGAL'' (project ID 855130) and from the German Excellence Strategy via the Heidelberg Cluster ``STRUCTURES'' (EXC 2181 - 390900948). In addition RSK is grateful for funding from the German Ministry for Economic Affairs and Climate Action in project ``MAINN'' (funding ID 50OO2206), and from DFG and ANR for project ``STARCLUSTERS'' (funding ID KL 1358/22-1). AR and LR gratefully
acknowledge funding from the DFG through an Emmy Noether Research Group
(grant number CH2137/1-1). 
\end{acknowledgements}

\footnotesize{
\bibliographystyle{aa}
\bibliography{Ref}
}

\begin{appendix}
\label{S:appendix_summary}

\onecolumn
\setlength{\tabcolsep}{3pt}
\section{Timescale across galaxies}
\begin{longtable}{lcccccccccc}
\caption{PHANGS-JWST Galaxy Sample}\\
\hline
\label{T:big_table}
Galaxy & RA [deg] & Dec [deg] & $\gamma$ & $N_0$ & $\log\,M_0$ [\msun] & $ T_{\rm dep}$ [Gyr] & $T_{\rm ff}$ [Myr] & $T_{\rm orb}$ [Myr] & $T_{\rm shear}$ [Myr] & $T_{\rm age}(m)$ [Myr] \\
\hline
\endfirsthead
\caption{continued.}\\
\hline
Galaxy & RA [deg] & Dec [deg] & $\gamma$ & $N_0$ & $\log\,M_0$ [\msun] & $T_{\rm dep}$ [Gyr] & $T_{\rm ff}$ [Myr] & $T_{\rm orb}$ [Myr] & $T_{\rm shear}$ [Myr] & $T_{\rm age}(m)$ [Myr] \\
\hline
\endhead
\hline
\endfoot
IC5273 & 344.86 & -37.70 & $-1.31^{+0.05}_{-0.07}$ & $224.27^{+90.36}_{-76.07}$ & $6.52^{+0.05}_{-0.05}$ &  & $12.84^{+4.20}_{-3.36}$ & $137.43^{+39.66}_{-35.39}$ & $76.70^{+6.90}_{-3.12}$ & $12.98^{+7.10}_{-4.39}$ \\
IC5332 & 353.61 & -36.10 & $-1.33^{+0.08}_{-0.12}$ & $140.40^{+90.52}_{-56.96}$ & $5.58^{+0.04}_{-0.04}$ & $1.55^{+4.11}_{-1.02}$ & $29.87^{+3.98}_{-5.14}$ &  &  & $12.40^{+6.63}_{-4.94}$ \\
NGC0628 & 24.17 & 15.78 & $-1.57^{+0.07}_{-0.07}$ & $72.47^{+31.15}_{-23.15}$ & $6.34^{+0.08}_{-0.08}$ & $2.28^{+4.06}_{-1.36}$ & $19.85^{+7.28}_{-5.40}$ & $149.02^{+56.51}_{-42.81}$ & $59.30^{+16.28}_{-12.93}$ & $19.19^{+13.50}_{-7.57}$ \\
NGC1087 & 41.60 & -0.50 & $-1.49^{+0.03}_{-0.05}$ & $167.09^{+29.16}_{-32.92}$ & $6.69^{+0.04}_{-0.03}$ & $3.85^{+10.27}_{-2.25}$ & $11.47^{+7.45}_{-3.70}$ & $129.72^{+48.47}_{-48.91}$ & $45.22^{+16.03}_{-15.17}$ & $26.29^{+10.88}_{-7.57}$ \\
NGC1097 & 41.58 & -30.28 & $-1.76^{+0.04}_{-0.03}$ & $32.94^{+11.04}_{-8.48}$ & $6.55^{+0.11}_{-0.10}$ &  & $24.60^{+8.54}_{-7.73}$ & $122.69^{+49.17}_{-48.07}$ & $39.58^{+15.47}_{-15.02}$ & $23.96^{+13.89}_{-8.90}$ \\
NGC1300 & 49.92 & -19.41 & $-1.74^{+0.03}_{-0.05}$ & $77.36^{+19.57}_{-22.49}$ & $6.43^{+0.07}_{-0.07}$ & $4.39^{+11.32}_{-2.96}$ & $22.91^{+8.39}_{-7.73}$ & $254.17^{+36.41}_{-34.09}$ & $108.79^{+7.59}_{-3.68}$ & $23.24^{+12.48}_{-9.08}$ \\
NGC1365 & 53.40 & -36.14 & $-1.83^{+0.03}_{-0.03}$ & $53.48^{+15.61}_{-8.58}$ & $6.24^{+0.05}_{-0.07}$ & $0.70^{+1.07}_{-0.41}$ & $30.11^{+11.22}_{-8.86}$ & $100.96^{+19.93}_{-19.62}$ & $50.91^{+3.20}_{-2.82}$ & $26.66^{+16.30}_{-10.92}$ \\
NGC1385 & 54.37 & -24.50 & $-1.60^{+0.05}_{-0.04}$ & $72.42^{+29.84}_{-20.24}$ & $6.78^{+0.13}_{-0.08}$ & $2.88^{+3.92}_{-1.66}$ & $12.07^{+7.04}_{-3.87}$ & $154.86^{+74.97}_{-55.38}$ & $45.76^{+19.01}_{-10.04}$ & $28.89^{+11.61}_{-7.87}$ \\
NGC1433 & 55.51 & -47.22 & $-1.67^{+0.03}_{-0.04}$ & $154.19^{+23.85}_{-23.59}$ & $6.04^{+0.05}_{-0.05}$ & $3.38^{+8.13}_{-2.19}$ & $27.51^{+5.72}_{-7.82}$ & $231.78^{+36.48}_{-51.69}$ & $77.55^{+6.30}_{-5.68}$ & $24.84^{+14.98}_{-10.67}$ \\
NGC1511 & 59.91 & -67.64 & $-1.75^{+0.08}_{-0.04}$ & $3.35^{+4.56}_{-0.98}$ & $7.39^{+0.15}_{-0.24}$ &  & $13.87^{+7.42}_{-4.51}$ & $125.27^{+19.48}_{-12.08}$ & $173.58^{+163.37}_{-59.81}$ & $32.37^{+10.68}_{-9.22}$ \\
NGC1512 & 60.98 & -43.35 & $-1.75^{+0.07}_{-0.05}$ & $64.51^{+39.39}_{-11.44}$ & $6.14^{+0.06}_{-0.09}$ & $2.72^{+6.07}_{-1.58}$ & $27.89^{+5.83}_{-7.83}$ & $225.69^{+32.10}_{-39.63}$ & $72.45^{+10.05}_{-12.35}$ & $21.56^{+11.60}_{-9.30}$ \\
NGC1546 & 63.65 & -56.06 & $-1.51^{+0.06}_{-0.08}$ & $41.41^{+17.57}_{-12.91}$ & $6.26^{+0.07}_{-0.09}$ &  & $22.20^{+9.51}_{-7.95}$ & $99.80^{+50.90}_{-41.42}$ & $26.41^{+1.82}_{-0.53}$ & $17.90^{+9.31}_{-6.93}$ \\
NGC1559 & 64.40 & -62.78 & $-1.56^{+0.05}_{-0.05}$ & $153.92^{+48.85}_{-40.99}$ & $6.73^{+0.06}_{-0.06}$ &  & $11.82^{+6.44}_{-3.42}$ & $196.75^{+52.00}_{-45.54}$ & $128.37^{+5.07}_{-13.13}$ & $26.66^{+10.81}_{-7.29}$ \\
NGC1566 & 65.00 & -54.94 & $-1.66^{+0.02}_{-0.03}$ & $158.58^{+22.39}_{-21.71}$ & $6.56^{+0.04}_{-0.03}$ & $3.37^{+7.35}_{-2.12}$ & $18.85^{+7.46}_{-6.24}$ & $161.44^{+76.99}_{-71.22}$ & $44.42^{+22.72}_{-17.39}$ & $25.09^{+14.18}_{-9.32}$ \\
NGC1637 & 70.37 & -2.86 & $-1.36^{+0.03}_{-0.05}$ & $193.75^{+47.71}_{-47.67}$ & $6.24^{+0.06}_{-0.03}$ &  & $18.37^{+6.16}_{-4.72}$ &  &  & $14.82^{+8.40}_{-5.91}$ \\
NGC1672 & 71.43 & -59.25 & $-1.63^{+0.02}_{-0.03}$ & $166.92^{+21.44}_{-19.79}$ & $6.60^{+0.03}_{-0.03}$ & $3.96^{+7.91}_{-2.39}$ & $15.67^{+5.52}_{-4.54}$ & $251.00^{+85.82}_{-93.50}$ & $83.25^{+26.77}_{-28.28}$ & $26.28^{+12.75}_{-8.56}$ \\
NGC1792 & 76.31 & -37.98 & $-1.53^{+0.03}_{-0.02}$ & $156.15^{+29.76}_{-19.39}$ & $6.79^{+0.05}_{-0.04}$ &  & $12.79^{+6.95}_{-4.09}$ & $190.82^{+70.61}_{-55.78}$ & $75.17^{+19.89}_{-15.09}$ & $23.53^{+9.64}_{-6.93}$ \\
NGC1809 & 75.52 & -69.57 & $-1.44^{+0.04}_{-0.04}$ & $98.21^{+24.01}_{-20.26}$ & $6.30^{+0.05}_{-0.05}$ &  & $17.57^{+9.33}_{-5.20}$ & $141.72^{+37.06}_{-37.36}$ & $67.45^{+2.04}_{-1.29}$ & $17.17^{+9.95}_{-5.74}$ \\
NGC2090 & 86.76 & -34.25 & $-1.59^{+0.11}_{-0.07}$ & $39.22^{+21.93}_{-9.87}$ & $5.97^{+0.02}_{-0.03}$ &  & $23.44^{+7.91}_{-5.96}$ & $64.75^{+31.28}_{-22.83}$ & $26.72^{+5.97}_{-1.31}$ & $12.49^{+9.47}_{-5.49}$ \\
NGC2283 & 101.47 & -18.21 & $-1.43^{+0.02}_{-0.06}$ & $164.18^{+27.09}_{-44.01}$ & $6.46^{+0.05}_{-0.05}$ &  & $15.66^{+7.33}_{-5.42}$ & $151.89^{+64.71}_{-61.12}$ & $51.78^{+19.50}_{-12.92}$ & $19.14^{+9.08}_{-7.65}$ \\
NGC2566 & 124.69 & -25.50 & $-1.70^{+0.03}_{-0.03}$ & $90.57^{+26.17}_{-20.49}$ & $6.44^{+0.08}_{-0.07}$ &  & $23.02^{+9.61}_{-8.22}$ & $165.13^{+37.92}_{-59.76}$ & $50.86^{+12.37}_{-17.51}$ & $23.85^{+11.67}_{-8.92}$ \\
NGC2775 & 137.58 & 7.04 & $-1.39^{+0.05}_{-0.05}$ & $640.21^{+148.45}_{-117.49}$ & $5.71^{+0.01}_{-0.02}$ &  & $29.65^{+5.28}_{-5.90}$ & $97.65^{+16.31}_{-12.34}$ & $49.80^{+2.01}_{-1.57}$ & $13.89^{+8.34}_{-5.88}$ \\
NGC2835 & 139.47 & -22.35 & $-1.54^{+0.04}_{-0.07}$ & $82.23^{+14.77}_{-26.11}$ & $6.35^{+0.06}_{-0.03}$ & $2.36^{+2.81}_{-1.38}$ & $20.10^{+7.11}_{-5.76}$ & $119.69^{+29.47}_{-39.14}$ & $44.80^{+9.32}_{-8.71}$ & $20.72^{+12.85}_{-7.84}$ \\
NGC2903 & 143.04 & 21.50 & $-1.39^{+0.03}_{-0.05}$ & $279.53^{+61.80}_{-87.03}$ & $6.31^{+0.05}_{-0.06}$ &  & $18.88^{+8.84}_{-5.73}$ & $133.45^{+27.81}_{-33.79}$ & $85.45^{+2.51}_{-7.90}$ & $15.29^{+8.02}_{-5.78}$ \\
NGC2997 & 146.41 & -31.19 & $-1.54^{+0.02}_{-0.03}$ & $328.38^{+44.85}_{-54.21}$ & $6.55^{+0.04}_{-0.03}$ &  & $17.23^{+7.32}_{-5.78}$ & $201.25^{+61.56}_{-79.05}$ & $65.20^{+19.19}_{-24.44}$ & $23.92^{+11.76}_{-7.73}$ \\
NGC3059 & 147.53 & -73.92 & $-1.55^{+0.03}_{-0.03}$ & $222.25^{+36.66}_{-34.42}$ & $6.56^{+0.03}_{-0.03}$ &  & $15.32^{+7.35}_{-4.61}$ & $220.37^{+60.64}_{-71.05}$ & $81.53^{+16.90}_{-12.36}$ & $23.98^{+12.04}_{-8.32}$ \\
NGC3137 & 152.28 & -29.06 & $-1.27^{+0.05}_{-0.09}$ & $218.14^{+92.77}_{-86.56}$ & $5.88^{+0.05}_{-0.02}$ &  & $26.18^{+11.83}_{-7.25}$ & $150.92^{+90.96}_{-47.13}$ & $82.49^{+1.99}_{-7.19}$ & $11.71^{+5.57}_{-4.62}$ \\
NGC3239 & 156.27 & 17.16 & $-1.32^{+0.13}_{-0.09}$ & $38.11^{+52.89}_{-19.98}$ & $6.07^{+0.16}_{-0.12}$ &  & $22.96^{+9.17}_{-7.36}$ &  &  & $17.48^{+8.10}_{-7.42}$ \\
NGC3351 & 160.99 & 11.70 & $-1.49^{+0.09}_{-0.08}$ & $127.19^{+67.17}_{-47.31}$ & $5.94^{+0.08}_{-0.07}$ & $1.76^{+3.20}_{-0.99}$ & $28.08^{+5.70}_{-5.94}$ & $97.82^{+14.05}_{-17.52}$ & $35.53^{+4.25}_{-5.18}$ & $16.07^{+9.54}_{-6.98}$ \\
NGC3507 & 165.86 & 18.14 & $-1.59^{+0.02}_{-0.04}$ & $204.92^{+32.98}_{-38.45}$ & $6.31^{+0.04}_{-0.04}$ &  & $22.76^{+6.80}_{-6.81}$ & $181.02^{+56.16}_{-52.14}$ & $68.29^{+9.89}_{-5.23}$ & $20.10^{+13.12}_{-7.69}$ \\
NGC3511 & 165.85 & -23.09 & $-1.33^{+0.06}_{-0.03}$ & $177.29^{+65.12}_{-38.46}$ & $6.47^{+0.05}_{-0.05}$ &  & $16.54^{+8.31}_{-5.01}$ & $180.39^{+83.04}_{-73.68}$ & $68.95^{+24.36}_{-19.31}$ & $17.46^{+7.24}_{-5.41}$ \\
NGC3521 & 166.45 & -0.03 & $-1.40^{+0.04}_{-0.04}$ & $424.02^{+92.50}_{-71.22}$ & $6.56^{+0.03}_{-0.03}$ &  & $13.08^{+8.35}_{-4.36}$ & $124.00^{+52.43}_{-36.32}$ & $47.70^{+6.65}_{-1.18}$ & $22.23^{+9.13}_{-6.44}$ \\
NGC3596 & 168.78 & 14.79 & $-1.41^{+0.05}_{-0.07}$ & $61.20^{+27.92}_{-25.29}$ & $6.34^{+0.11}_{-0.10}$ &  & $17.70^{+6.94}_{-4.62}$ & $71.91^{+24.13}_{-27.45}$ & $29.84^{+7.56}_{-5.13}$ & $21.38^{+11.73}_{-8.00}$ \\
NGC3621 & 169.57 & -32.81 & $-1.48^{+0.05}_{-0.04}$ & $82.68^{+26.81}_{-17.79}$ & $6.74^{+0.03}_{-0.05}$ &  & $11.42^{+5.92}_{-3.53}$ & $131.28^{+40.52}_{-36.58}$ & $65.89^{+6.12}_{-2.33}$ & $24.70^{+10.82}_{-6.09}$ \\
NGC3626 & 170.02 & 18.36 & $-1.69^{+0.29}_{-0.12}$ & $4.20^{+36.50}_{-2.55}$ & $6.48^{+0.23}_{-0.59}$ &  & $25.23^{+7.12}_{-5.88}$ & $26.94^{+7.29}_{-5.55}$ & $9.71^{+2.09}_{-1.55}$ & $11.77^{+7.39}_{-5.09}$ \\
NGC3627 & 170.06 & 12.99 & $-1.86^{+0.20}_{-0.05}$ & $6.02^{+42.13}_{-2.19}$ & $7.34^{+0.24}_{-0.52}$ & $1.59^{+4.37}_{-0.88}$ & $15.12^{+6.34}_{-4.53}$ & $109.59^{+51.32}_{-37.66}$ & $40.80^{+5.88}_{-1.04}$ & $29.06^{+14.03}_{-10.94}$ \\
NGC4254 & 184.71 & 14.42 & $-1.50^{+0.02}_{-0.03}$ & $221.58^{+38.20}_{-39.08}$ & $6.80^{+0.04}_{-0.04}$ & $4.36^{+6.66}_{-2.85}$ & $10.65^{+4.55}_{-3.05}$ & $163.06^{+77.90}_{-52.77}$ & $57.82^{+24.13}_{-17.39}$ & $25.01^{+11.31}_{-7.01}$ \\
NGC4298 & 185.39 & 14.61 & $-1.51^{+0.07}_{-0.07}$ & $84.75^{+41.83}_{-22.80}$ & $6.23^{+0.06}_{-0.06}$ &  & $20.44^{+6.67}_{-5.16}$ & $143.48^{+45.31}_{-41.53}$ & $62.10^{+12.50}_{-7.38}$ & $20.03^{+10.13}_{-6.55}$ \\
NGC4303 & 185.48 & 4.47 & $-1.58^{+0.02}_{-0.01}$ & $246.10^{+23.97}_{-14.02}$ & $6.67^{+0.03}_{-0.02}$ & $2.45^{+5.04}_{-1.46}$ & $14.21^{+7.96}_{-4.78}$ & $135.35^{+34.71}_{-56.46}$ & $43.74^{+11.46}_{-16.96}$ & $24.29^{+12.94}_{-7.97}$ \\
NGC4321 & 185.73 & 15.82 & $-1.57^{+0.03}_{-0.03}$ & $192.90^{+54.59}_{-21.26}$ & $6.43^{+0.06}_{-0.04}$ & $1.99^{+2.77}_{-1.09}$ & $19.95^{+6.83}_{-5.52}$ & $180.56^{+57.83}_{-62.88}$ & $65.78^{+17.30}_{-19.62}$ & $19.69^{+13.35}_{-7.28}$ \\
NGC4424 & 186.80 & 9.42 & $-1.84^{+0.10}_{-0.07}$ & $1.54^{+2.96}_{-1.40}$ & $6.67^{+1.05}_{-0.33}$ &  & $21.47^{+7.48}_{-8.22}$ &  &  & $25.34^{+10.23}_{-6.85}$ \\
NGC4457 & 187.25 & 3.57 & $-2.09^{+0.14}_{-0.10}$ & $3.28^{+4.45}_{-2.91}$ & $6.12^{+0.71}_{-0.16}$ &  & $29.22^{+9.38}_{-13.06}$ & $39.60^{+19.02}_{-13.09}$ & $9.78^{+3.63}_{-2.16}$ & $20.06^{+9.43}_{-7.61}$ \\
NGC4496A & 187.91 & 3.94 & $-1.39^{+0.05}_{-0.04}$ & $143.77^{+51.13}_{-33.82}$ & $6.32^{+0.06}_{-0.06}$ &  & $20.79^{+8.59}_{-8.52}$ & $240.44^{+23.60}_{-22.27}$ & $312.89^{+491.97}_{-87.94}$ & $18.92^{+9.43}_{-7.34}$ \\
NGC4535 & 188.58 & 8.20 & $-1.69^{+0.04}_{-0.10}$ & $48.70^{+16.34}_{-21.89}$ & $6.52^{+0.19}_{-0.09}$ & $3.68^{+5.84}_{-2.73}$ & $22.92^{+8.10}_{-7.59}$ & $171.00^{+47.96}_{-54.34}$ & $56.52^{+14.86}_{-16.78}$ & $21.08^{+12.04}_{-8.22}$ \\
NGC4536 & 188.61 & 2.19 & $-1.51^{+0.04}_{-0.06}$ & $128.80^{+37.43}_{-35.73}$ & $6.36^{+0.07}_{-0.07}$ &  & $23.33^{+10.60}_{-8.07}$ & $178.63^{+132.73}_{-81.98}$ & $56.92^{+42.22}_{-26.03}$ & $24.75^{+11.13}_{-9.26}$ \\
NGC4540 & 188.71 & 15.55 & $-1.51^{+0.07}_{-0.09}$ & $41.92^{+21.31}_{-17.61}$ & $6.29^{+0.10}_{-0.08}$ &  & $19.48^{+7.93}_{-5.34}$ & $122.72^{+36.40}_{-37.11}$ & $66.35^{+4.97}_{-6.84}$ & $18.45^{+12.56}_{-6.93}$ \\
NGC4548 & 188.86 & 14.50 & $-1.49^{+0.07}_{-0.06}$ & $261.73^{+119.58}_{-59.96}$ & $5.95^{+0.05}_{-0.05}$ &  & $28.32^{+5.89}_{-8.30}$ & $169.03^{+31.80}_{-59.02}$ & $46.44^{+9.89}_{-17.75}$ & $14.07^{+8.84}_{-5.62}$ \\
NGC4569 & 189.21 & 13.16 & $-1.56^{+0.09}_{-0.30}$ & $43.37^{+39.93}_{-40.24}$ & $6.10^{+0.78}_{-0.16}$ &  & $28.89^{+4.92}_{-8.06}$ & $198.20^{+23.27}_{-42.86}$ & $150.64^{+8.08}_{-24.34}$ & $17.36^{+11.03}_{-5.95}$ \\
NGC4571 & 189.23 & 14.22 & $-1.43^{+0.07}_{-0.07}$ & $271.40^{+128.03}_{-72.69}$ & $5.82^{+0.03}_{-0.04}$ &  & $28.22^{+5.27}_{-6.22}$ & $144.95^{+56.73}_{-59.46}$ & $57.54^{+18.03}_{-16.02}$ & $15.60^{+7.93}_{-6.40}$ \\
NGC4579 & 189.43 & 11.82 & $-1.69^{+0.04}_{-0.04}$ & $124.52^{+31.84}_{-31.66}$ & $6.05^{+0.06}_{-0.03}$ &  & $27.98^{+6.04}_{-6.
43}$ & $150.21^{+42.40}_{-41.95}$ & $47.17^{+10.29}_{-8.03}$ & $20.73^{+13.59}_{-8.67}$ \\
NGC4654 & 190.99 & 13.13 & $-1.57^{+0.02}_{-0.02}$ & $213.03^{+37.59}_{-25.67}$ & $6.54^{+0.03}_{-0.03}$ &  & $19.34^{+8.61}_{-6.19}$ & $252.62^{+77.66}_{-65.00}$ & $112.86^{+11.01}_{-3.22}$ & $24.57^{+11.66}_{-7.81}$ \\
NGC4689 & 191.94 & 13.76 & $-1.37^{+0.04}_{-0.07}$ & $270.06^{+71.63}_{-80.21}$ & $5.99^{+0.03}_{-0.04}$ &  & $23.86^{+6.62}_{-5.77}$ & $147.87^{+38.05}_{-48.80}$ & $60.66^{+10.44}_{-11.56}$ & $13.60^{+9.15}_{-5.74}$ \\
NGC4694 & 192.06 & 10.98 & $-1.41^{+0.13}_{-0.34}$ & $31.55^{+30.65}_{-30.61}$ & $5.76^{+1.17}_{-0.14}$ &  & $22.38^{+10.15}_{-5.76}$ &  &  & $16.84^{+9.79}_{-6.51}$ \\
NGC4731 & 192.76 & -6.39 & $-1.37^{+0.12}_{-0.04}$ & $67.39^{+77.84}_{-20.50}$ & $6.30^{+0.09}_{-0.11}$ &  & $16.78^{+10.14}_{-5.82}$ &  &  & $16.36^{+11.45}_{-6.66}$ \\
NGC4781 & 193.60 & -10.54 & $-1.47^{+0.05}_{-0.04}$ & $79.20^{+21.15}_{-18.23}$ & $6.69^{+0.05}_{-0.04}$ &  & $10.65^{+5.13}_{-2.66}$ & $117.00^{+45.78}_{-39.59}$ & $49.07^{+14.25}_{-9.56}$ & $23.37^{+9.98}_{-7.79}$ \\
NGC4826 & 194.18 & 21.68 & $-1.51^{+0.22}_{-0.14}$ & $6.43^{+14.40}_{-4.24}$ & $6.38^{+0.44}_{-0.31}$ &  & $16.90^{+16.36}_{-5.12}$ & $27.58^{+7.17}_{-7.05}$ & $9.04^{+2.28}_{-2.27}$ & $16.07^{+7.77}_{-5.48}$ \\
NGC4941 & 196.05 & -5.55 & $-1.14$ & $873.37$ & $5.59$ &  & $32.00^{+6.78}_{-7.01}$ & $126.21^{+29.37}_{-29.54}$ & $62.16^{+4.44}_{-4.56}$ & $3.77^{+1.34}_{-1.75}$ \\
NGC4951 & 196.28 & -6.49 & $-1.28^{+0.07}_{-0.61}$ & $124.75^{+66.33}_{-123.71}$ & $6.08^{+1.26}_{-0.08}$ &  & $18.78^{+9.33}_{-5.08}$ & $87.19^{+61.02}_{-27.26}$ & $37.50^{+9.53}_{-0.91}$ & $13.35^{+5.40}_{-4.80}$ \\
NGC5042 & 198.88 & -23.98 & $-1.40^{+0.03}_{-0.06}$ & $182.20^{+33.88}_{-49.18}$ & $6.16^{+0.04}_{-0.04}$ &  & $23.36^{+8.47}_{-6.53}$ & $161.23^{+61.21}_{-50.05}$ & $67.12^{+5.42}_{-2.58}$ & $16.26^{+9.53}_{-6.23}$ \\
NGC5068 & 199.73 & -21.04 & $-1.64^{+0.24}_{-0.19}$ & $13.09^{+45.41}_{-6.67}$ & $6.62^{+0.09}_{-0.33}$ & $1.96^{+3.19}_{-1.22}$ & $18.31^{+6.66}_{-6.27}$ & $219.42^{+65.71}_{-64.52}$ & $107.49^{+9.40}_{-6.23}$ & $23.74^{+17.63}_{-9.93}$ \\
NGC5134 & 201.33 & -21.13 & $-1.50^{+0.07}_{-0.05}$ & $105.89^{+51.22}_{-25.24}$ & $6.16^{+0.09}_{-0.07}$ &  & $22.69^{+7.40}_{-6.86}$ & $141.27^{+44.77}_{-38.34}$ & $75.30^{+8.55}_{-8.50}$ & $15.48^{+11.32}_{-7.08}$ \\
NGC5248 & 204.38 & 8.89 & $-1.63^{+0.05}_{-0.03}$ & $68.66^{+21.58}_{-17.06}$ & $6.54^{+0.08}_{-0.08}$ &  & $17.88^{+7.73}_{-5.21}$ & $136.84^{+62.08}_{-34.93}$ & $57.65^{+6.87}_{-2.07}$ & $23.43^{+11.58}_{-7.73}$ \\
NGC5643 & 218.17 & -44.17 & $-1.46^{+0.03}_{-0.03}$ & $236.02^{+49.93}_{-34.84}$ & $6.53^{+0.03}_{-0.04}$ &  & $16.04^{+6.82}_{-5.22}$ & $136.14^{+43.92}_{-51.99}$ & $46.78^{+13.43}_{-16.63}$ & $18.21^{+11.01}_{-6.60}$ \\
NGC6300 & 259.25 & -62.82 & $-1.58^{+0.05}_{-0.07}$ & $65.40^{+23.23}_{-23.54}$ & $6.44^{+0.10}_{-0.06}$ &  & $19.43^{+8.33}_{-5.68}$ & $102.79^{+36.28}_{-38.68}$ & $32.36^{+11.53}_{-12.26}$ & $22.09^{+12.42}_{-8.67}$ \\
NGC7456 & 345.54 & -39.57 & $-1.32^{+0.11}_{-0.06}$ & $112.45^{+90.16}_{-29.56}$ & $5.79^{+0.04}_{-0.06}$ &  & $28.76^{+10.14}_{-8.27}$ & $197.80^{+45.55}_{-47.65}$ & $132.44^{+3.28}_{-12.36}$ & $14.90^{+6.87}_{-5.50}$ \\
NGC7496 & 347.45 & -43.43 & $-1.55^{+0.03}_{-0.03}$ & $125.09^{+27.77}_{-25.05}$ & $6.56^{+0.07}_{-0.05}$ & $7.25^{+21.08}_{-4.99}$ & $16.38^{+6.39}_{-5.02}$ & $207.10^{+69.40}_{-77.18}$ & $72.74^{+20.92}_{-21.97}$ & $25.93^{+11.54}_{-8.75}$ \\
\end{longtable}
\tablefoot{\textit{Global properties of the galaxies:} Right Ascension (RA), Declination (Dec). \textit{Truncated power law fit parameters:} $\gamma$, $N_{0}$, and $M_{0}$. \textit{Cloud timescales:} Median cloud \tdep, \tff, \torb , \tshear, and $T_{\rm age}(m)$. The 84th - 50th and 50th - 16th percentiles are shown in superscript and subscript, respectively. Square brackets refer to upper limit estimates.}
\setlength{\tabcolsep}{6pt}

\onecolumn
\section{The completeness limit} \label{a:completness}

\begin{figure*}[h]
    \centering
    \includegraphics[width=0.98\textwidth]{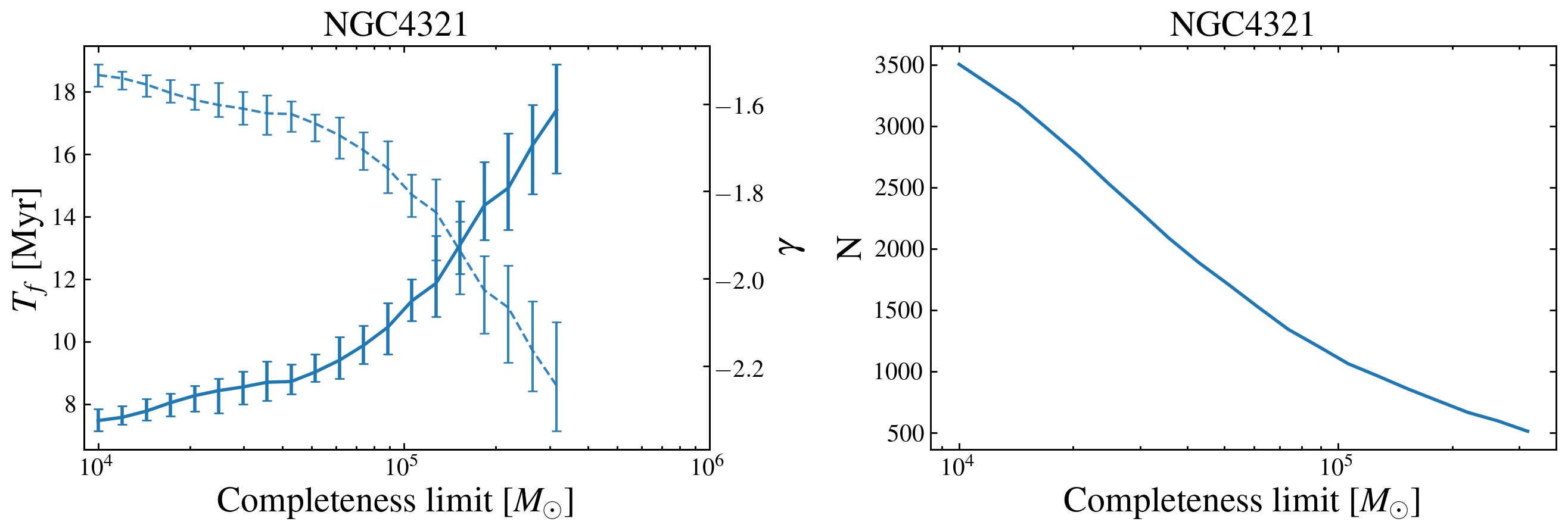}
    \caption{\textit{Left:} The variation of the $T_{\rm f}$ (solid line) and $\gamma$ (dashed line) as a function of the completeness limit in NGC4321. Error bars represent the fit scatter from running 100 random sampling bootstrap iterations on each completeness limit value. \textit{Right:} Number of clouds (N) as a function of the completeness limit.}
    \label{fig:Completeness}
\end{figure*}

In this section, we test the impact of the completeness limit on $T_{\rm f}$ (see Fig.~\ref{fig:Completeness}). Generally, upon increasing the completness limit from $10^{4}\, \msun$ to $2\times10^{5}\, \msun$, we find that $T_{\rm f}$ increases too. However, the number of clouds in the fit decreases, and the error on $T_{\rm f}$ also increases. While this variation could be purely due to statistics (e.g., sample size, incompleteness), it also highlights that the slope $\gamma$ of the mass spectrum is not universal at all masses (see also \citealt{Colombo_2014}).  

\section{The truncation limit} \label{a:trunc}
\begin{figure*}[h]
    \centering
    \includegraphics[width=0.45\textwidth]{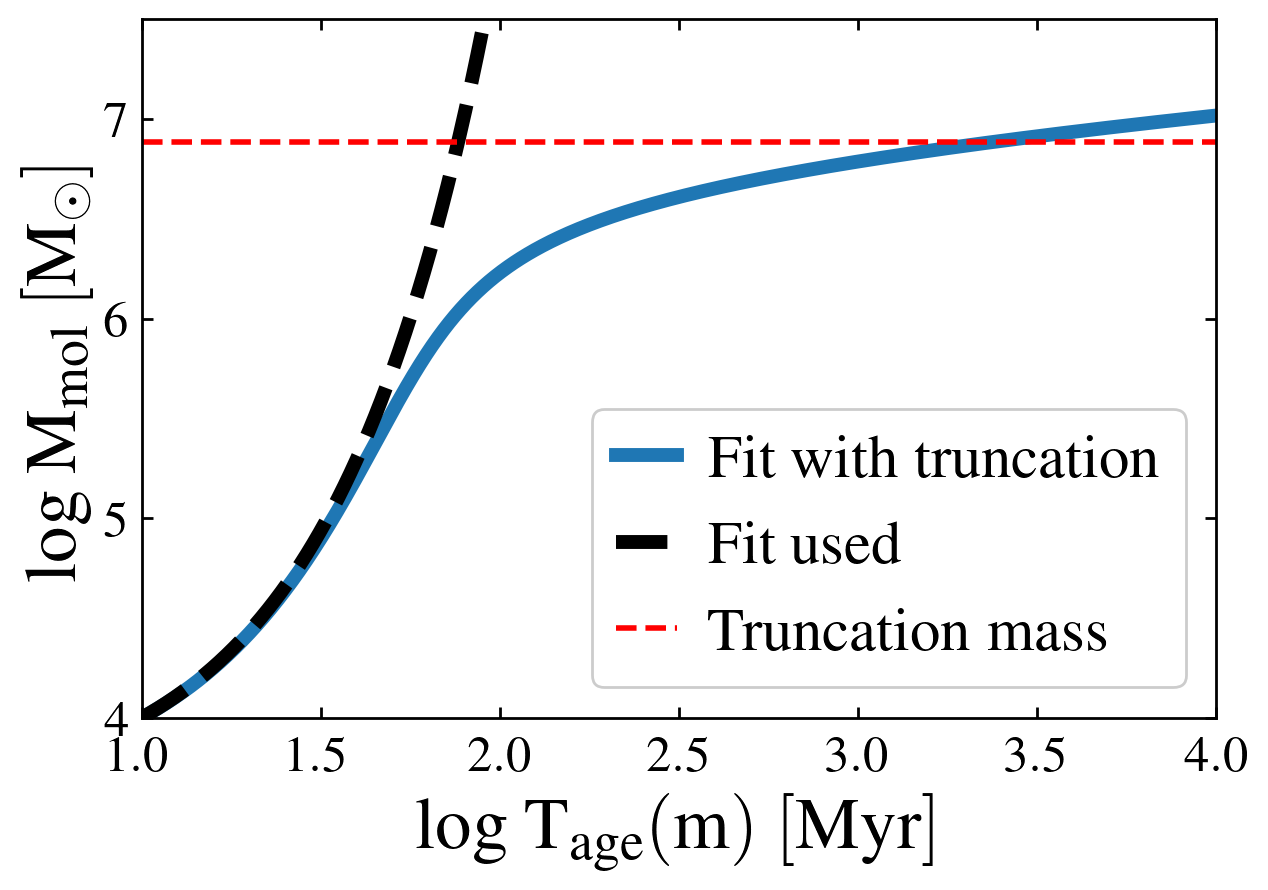}
    \caption{\textit{Left:} Individual GMC \Mmol\ as a function of $T_{\rm age}(m)$ following Eq.~\ref{eq:age_integral}, where the black dashed line shows cloud exponential growth without a truncation, and the blue curve shows the cloud growth with a truncation mass of $7.7\times 10^{6}\,\msun$ (calculated for a shell of 100 pc radius and HI density of $10 \rm\, cm^{-3}$; see expanding shell argument in \citealt{Kobayashi2017}). The truncation mass is shown with the dashed red line.}
    \label{fig:Truncation}
\end{figure*}

Our analysis follows that clouds follow a constant $T_{\rm f}$ per galaxy or environment for all clouds. However, as explained in \cite{Kobayashi2017}, the mass-independent formulations overestimate the growth rate of very massive GMCs whose mass is comparable to the mass of a shell swept up by an expanding bubble. Thus, a truncation can be introduced in the form of $T_{\rm f}$(m) $ =  T_{\rm f}\big[(1 + m/m_{\rm trunc})^{\beta}\big]$ as explained in Section~\ref{S:timescales}. Upon Taylor expansion, $T_{\rm f}$(m) = $T_{\rm f}\big[(1 + \beta m/m_{\rm trunc})\big]$, and $m_{\rm crit} = m_{\rm trunc}/\beta$, where $m_{\rm crit}$ is the typical maximum GMC mass that can be created in the \cite{Inutsuka_2015} evolution scenario. In Fig.~\ref{fig:Truncation}, we show how the age of a GMC varies when we add a truncation, where deviations start to occur at the more massive end of the mass spectrum ($\Mmol \gtrsim  10^{6}\, \msun $), and thus we opt to fit without this truncation in our presented analysis. However, it is worth noting that the deviation is solely connected to the maximum GMC mass that can be swept up from a supernova remnant. Assuming multiple supernovae occur, or that the radius of the swept-up shell is greater than 100 pc, the truncation fit becomes closer to the GMC exponential growth fit that we used.

\onecolumn
\section{The CO-to-H$_{2}$ conversion factor} \label{a:convfact}

\renewcommand{\arraystretch}{1.5}
\setlength{\tabcolsep}{2pt}
\begin{table*}[h]
\caption{Comparison of truncated power-law fit parameters between SL24 (variable $\alpha_{\rm CO}$; \citealt{schinnerer2024}) and cst (constant $\alpha_{\rm CO}$), where the percentage differences are computed as $(\mathrm{SL24}-\mathrm{cst})/\mathrm{cst}$.}
\centering
\begin{tabular}{ccccccc}
\hline
Env. &
$\gamma_{\rm SL24}$ &
$\gamma_{\rm cst}$ &
$\Delta\gamma$ [\%] &
$T_{\rm f,\, SL24}$ [Myr] &
$T_{\rm f,\, cst}$ [Myr] &
$\Delta T_{\rm f}$ [\%] \\
\hline

Global &
$-1.51^{+0.13}_{-0.18}$ &
$-1.59^{+0.15}_{-0.30}$ &
$-3.0^{+9.2}_{-10.9}$ &
$7.34^{+2.23}_{-2.11}$ &
$7.78^{+3.43}_{-2.14}$ &
$-8.8^{+27.8}_{-26.4}$ \\

Center &
$-1.34^{+0.05}_{-0.12}$ &
$-1.40^{+0.12}_{-0.11}$ &
$-1.4^{+4.8}_{-11.7}$ &
$5.09^{+1.91}_{-1.79}$ &
$6.18^{+1.86}_{-2.15}$ &
$-5.2^{+16.4}_{-32.0}$ \\

Bar &
$-1.61^{+0.21}_{-0.14}$ &
$-1.60^{+0.23}_{-0.13}$ &
$-0.7^{+7.5}_{-10.4}$ &
$8.44^{+2.59}_{-2.94}$ &
$8.59^{+2.82}_{-2.53}$ &
$-1.8^{+18.8}_{-27.2}$ \\

S-Arm &
$-1.48^{+0.09}_{-0.10}$ &
$-1.50^{+0.11}_{-0.15}$ &
$-2.5^{+4.8}_{-6.3}$ &
$6.79^{+1.37}_{-1.53}$ &
$7.74^{+2.19}_{-2.55}$ &
$-7.7^{+15.4}_{-15.9}$ \\

I-Arm &
$-1.59^{+0.13}_{-0.07}$ &
$-1.63^{+0.08}_{-0.11}$ &
$-1.6^{+11.3}_{-7.7}$ &
$8.06^{+1.37}_{-1.72}$ &
$8.08^{+2.48}_{-1.80}$ &
$-3.8^{+35.2}_{-19.6}$ \\

Disk &
$-1.49^{+0.13}_{-0.10}$ &
$-1.52^{+0.15}_{-0.13}$ &
$-1.5^{+6.6}_{-6.4}$ &
$6.68^{+1.92}_{-1.97}$ &
$6.98^{+2.72}_{-2.79}$ &
$-4.4^{+24.3}_{-21.8}$ \\

\hline
\end{tabular}
\tablefoot{
$T_{\rm f}$ and $\gamma$ values are medians of one unique formation timescale per galaxy and environment, 
with uncertainties derived from 100 bootstrap resamples.
}
\label{tab:sl24_b13_full_comparison}
\end{table*}

In our analysis, we used the \cite{schinnerer2024} prescription of the \aco\ conversion factor (see Sect.~\ref{S:MC_Props}). However, multiple prescriptions exist and might bias the timescale estimates through the change of the slope $\gamma$ of the GMC mass spectra. Therefore, to test if our analysis holds with another definition, we use a the constant Galactic $\alpha_{\rm CO(2-1)} = \frac{4.35}{0.65} = 6.69\, \rm \msunperpcsq (K\,km\,s^{-1})^{-1}$, where $R_{21}= 0.65$ is based on \cite{Leroy_2013} and \cite{denbrok_2020}, measured at kpc scales, and $\alpha_{\rm CO(1-0)} = 4.35~\rm \msunperpcsq (K\,km\,s^{-1})^{-1}$ is the standard Galactic value at solar metallicity (i.e., \citealt{Bolatto_2013}). Generally, as depicted in Table~\ref{tab:sl24_b13_full_comparison}, we find that $\gamma$ and $T_{\rm f}$ only vary by less than $10\,\%$, on average, upon adopting a different \aco\ prescription. Therefore, the results presented in the analysis are robust to the choice of \aco.

\section{Additional Plots}

\begin{figure}[h]
    \centering
    \includegraphics[width=0.49\textwidth]{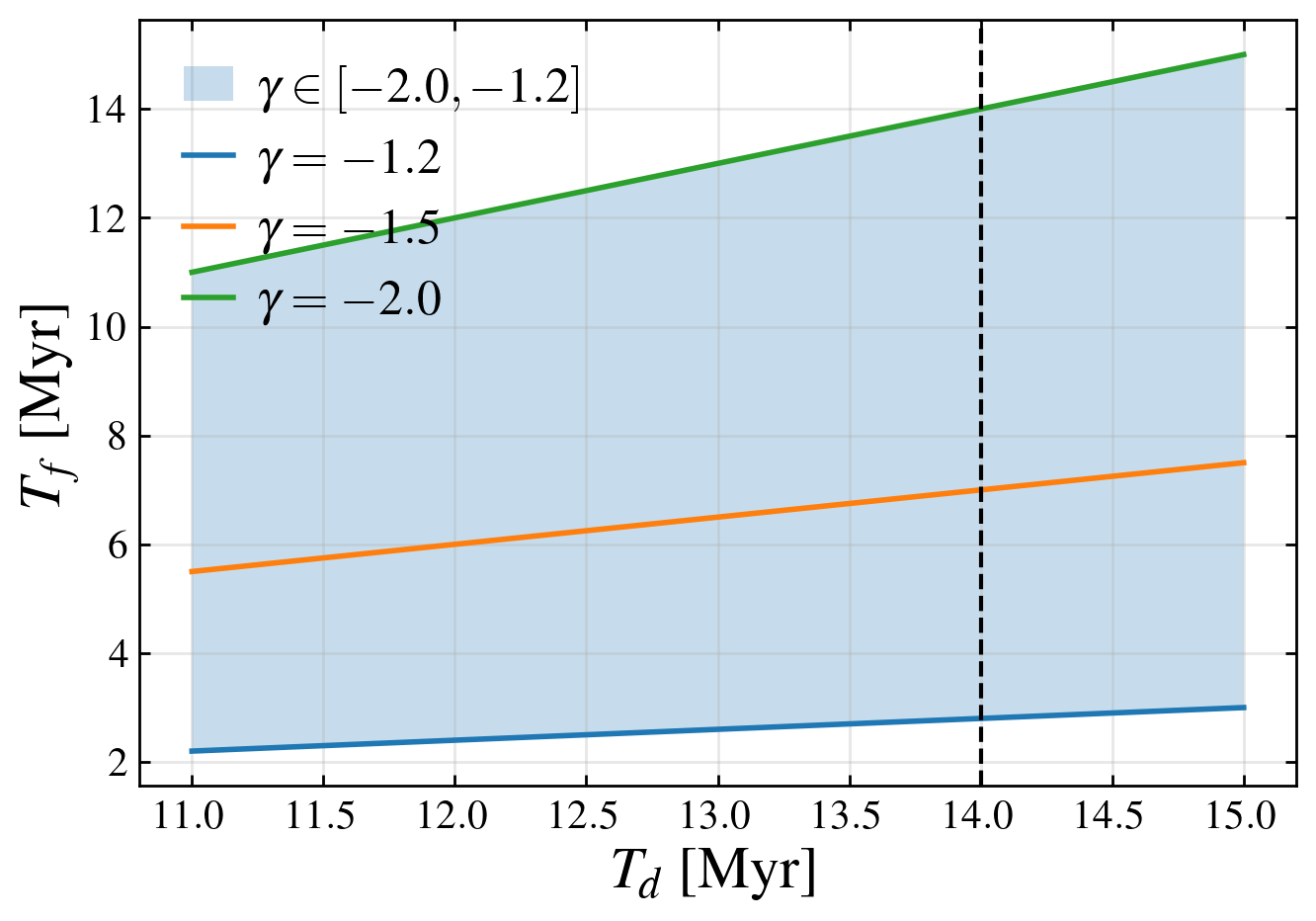}
    \caption{$T_{\rm f}$ as a function of $T_{d}$ for different $\gamma$ indexes. The vertical dashed black line is the adopted value of $T_{d} = 14$~Myr in our analysis. We vary $\gamma$ from $\rm -1.2$ to $\rm -2.0$, which are the typical values across the galaxies, and note that $T_{\rm f}$ varies according to $T_{\rm f} = -(\gamma + 1)\times T_{d}$. Thus, higher values of $T_{d}$ lead to more variation between $T_{\rm f}$ across the galaxies as depicted by the blue shaded regions.}
    \label{fig:Tf_Td}
\end{figure}

\end{appendix}

\end{document}